\documentclass[12pt]{article}
\usepackage{ascmac}
\usepackage{ulem}
\usepackage{latexsym}
\usepackage[dvipdfmx]{graphicx}
\usepackage{amsmath}
\usepackage[top=3.5cm,bottom=3.5cm,left=2cm,right=2cm]{geometry}
\usepackage{pifont}          
\usepackage{amsmath,amssymb}
\usepackage{color}
\usepackage{geometry}
\usepackage{caption}
\usepackage{ulem}
\usepackage{cancel}
\newcommand{\nn}{\nonumber\\}

\newcommand{\h}{\hspace}
\newcommand{\be}{\begin{equation}}
\newcommand{\e}{\end{equation}}
\newcommand{\aln}[1]{\begin{align}#1\end{align}}

\begin{document}

\title{
\vbox{
\baselineskip 14pt
\hfill \hbox{\normalsize KEK-TH-2018
}} \vskip 1cm
\bf \Large  RG-improvement of the effective action with multiple mass scales
\vskip 0.5cm
}
\author{
Satoshi Iso$^{a,b}$\thanks{E-mail: \tt satoshi.iso(at)kek.jp}\h{1mm} 
and Kiyoharu~Kawana$^{a}$\thanks{E-mail: \tt kawana(at)post.kek.jp}
\bigskip\\
\it 
\normalsize
 $^a$ Theory Center, High Energy Accelerator Research Organization (KEK), \\
 \normalsize
\it  $^b$Graduate University for Advanced Studies (SOKENDAI),\\
\\
 \normalsize
\it Tsukuba, Ibaraki 305-0801, Japan \\
\smallskip
}
\date{\today}

\maketitle

\abstract{\normalsize
Improving the effective action by the renormalization group (RG) 
with several mass scales is an important problem in quantum field theories. 
A method  based on the decoupling theorem was proposed in \cite{Bando:1992wy} and systematically
improved \cite{Casas:1998cf} to take threshold effects into account. 
In this paper,  we apply the method to the Higgs-Yukawa model, 
including wave-function renormalizations, and  to 
a model with two real scalar fields $(\varphi, h)$. 
In the Higgs-Yukawa model, even at  one-loop level,
Feynman diagrams contain propagators with different mass scales 
and  decoupling scales must be chosen appropriately to absorb threshold corrections. 
On the other hand, in the two-scalar model, the mass matrix of the scalar fields
is a function of their field values  $(\varphi, h)$ and 
the resultant running couplings  obey different RGEs on a different point of the  field space. 
By solving the RGEs, 
we  can obtain the RG improved effective action in the whole region of the scalar fields.  
}
\newpage

\section{Introduction}

An effective field theory is a powerful tool to study physics at low energy scale. 
The standard model (SM) of particle physics is the most successful effective theory in nature 
that describes physics below TeV scale, from which we can infer physics at higher energy scales.
For example, by using the recently observed Higgs boson mass $\sim 125$GeV, 
the running quartic Higgs coupling $\lambda(\mu)$ is shown to decrease
 as the energy $\mu$ increases. 
It may either cross zero around $10^{11}$ GeV, inducing instability of the vacuum,  
or may asymptotically vanish around the Planck scale, which indicates a possibility that the SM
is safely interpolated up to the Planck scale without  instability \cite{Holthausen:2011aa,Bezrukov:2012sa,Degrassi:2012ry,Iso:2012jn,Buttazzo:2013uya,Kawana:2015tka,Hamada:2015fma}. 
The behavior of 
the running quartic coupling is quite sensitive to 
 the precise values of various SM parameters such as the top quark mass or the Higgs mass itself \cite{Alekhin:2012py,Moch:2014tta,Cortiana:2015rca}. 
The importance of  $\lambda(\mu)$ is of course related to the shape of 
the renormalization group (RG) improved effective potential $V(\phi) \sim \lambda(\mu(\phi)) \phi^4$. 
If there is only a single mass scale $\phi$, we can safely set 
the renormalization scale $\mu$ at $\mu=\phi$. However, if there are multiple scales
that are different from each other,  e.g. $M(\phi) \gg m(\phi)$, 
a careful analysis of the effective potential is necessary  since a naive choice $\mu=\phi$ 
generates a large logarithm $\log (M/m) \gg 1$ in the effective potential. 
Such a situation particularly arises  when  the SM is extended to contain multiple scalar fields \footnote{In the SM itself, 
since the coupling to the Higgs boson determines its mass, light particles are weakly coupled
to the Higgs boson and do not contribute much to the Higgs effective potential. 
Thus, though SM has multiple hierarchical mass scales, we can safely set the renormalization scale 
$\mu$ at the heaviest particle mass,
i.e. at the top quark mass; the problem of large logarithms is usually not a big issue. The problem becomes
important when we extend the SM to contain additional scalar fields that are coupled to the SM.}. 

Suppose that we have multiple scalar fields $\{ \phi_{a}, a=1,2, \cdots \}$
 and  that various particles acquire masses $M_i(\{ \phi_a \})$ through the vacuum expectation values of 
 these scalar fields. 
At  one-loop level in the mass independent scheme, the effective potential is calculated as
\aln{ V(\{ \phi_a^{} \})
&\equiv V^{(0)}(\{\phi_a^{}\})+\sum_i V_i^{(1)}(\{\phi_a^{}\})\nonumber
\\
&=V^{(0)}(\{\phi_a^{}\})+\sum_{i} (-1)^{2s_i^{}} n_i^{}\frac{M_i^4(\{\phi_a^{}\})}{64\pi^2}\ln\left(\frac{M_i^2(\{\phi_a^{}\})}{\tilde{\mu}_i^2}\right),
\label{eq:general one-loop potential}
}
where $V^{(0)}(\{ \phi_a^{} \})$ is the tree level potential, 
$s_i^{}$ and $n_i^{}$ represent the spin and the number of degrees of freedom respectively.
$M_i^{}(\{\phi_a^{}\})$ are their mass eigenvalues and 
$\tilde{\mu}_i^{2}=\mu^2e^{C_i^{}}$ are the renormalization scales 
with the scheme dependent constants\footnote{
For instance, in the $\overline{\text{MS}}$ scheme, $C_i$ are
$3/2$, $3/2$ and $5/6$ for scalars, fermions and gauge bosons respectively.}. 
If there exists only a single mass scale $M(\phi)$, we can eliminate the one-loop terms by choosing $\tilde\mu=M(\phi)$. 
Such a choice with the RG-improved couplings
corresponds to  resumming  leading-logarithms to all orders in 
perturbative calculations \cite{Coleman:1973jx}. 
In contrast, if there are several mass scales with very different values, e.g. 
$M(\{ \phi_a \})\gg m(\{ \phi_a \})$, a naive choice $\tilde\mu=M$ cannot remove all the large logarithms, 
and the logarithmic factor such as $\ln(M(\{ \phi_a \} )^2/m(\{ \phi_a \})^2)$ may invalidate 
the perturbative calculation.

There are two different but related approaches to handle the issue of multi mass scales: multi-scale renormalization \cite{Einhorn:1983fc,Ford:1996yc} and a decoupling method \cite{Bando:1992wy,Bando:1992np}. In the former approach, several independent renormalization scales $(\mu_1^{},\mu_2^{},\cdots )$ are introduced hoping that $\tilde{\mu}_i^{}$'s in Eq.(\ref{eq:general one-loop potential}) might be replaced by $\mu_i^{}$'s. 
If such replacements actually occur, 
all the logarithms can be absorbed by putting $\tilde\mu_i^{}=M_i^{}(\{\phi_a^{}\})$. 
However, as discussed in \cite{Ford:1996yc,Steele:2014dsa}, 
each RG scale $\mu_i$ produces a different 
renormalization group equation (RGE), 
and it is difficult to solve them 
keeping the integrability conditions $[{\cal{D}}_i^{},{\cal{D}}_j^{}]=0$. 
Furthermore the $\beta$-functions generically contain logarithms of the
ratio of different renormalization scales $\mu_i$ at higher loops. 
Thus the perturbative validity will be lost, if there are hierarchical mass scales, e.g.  $\mu_i \ll \mu_j$ for $i \neq j$.

On the other hand, the decoupling method is based on the decoupling theorem in field theories \cite{Appelquist:1974tg,Symanzik:1973vg}.
Below a decoupling scale, 
massive particles with mass $M$ can be integrated out and their effects 
are absorbed into the effective couplings 
and higher dimensional operators in the effective field theory of light particles, whose masses
are given by a single mass scale denoted by $m$. 
The radiative corrections in the effective field theory are then given by a single type of logarithms
$\ln (m^2/\mu^2)$, and consequently we can improve the effective potential by setting $\mu=m.$ 
The idea is given in \cite{Bando:1992wy,Bando:1992np} and applied to the Higgs-Yukawa model.
But in order to avoid large threshold corrections containing $\ln (M^2/m^2)$,
we need to carefully choose the decoupling scale.
A systematic procedure to handle such threshold corrections 
is proposed by \cite{Casas:1998cf}, in which the authors  choose the decoupling scale 
so that the large logarithms  $\ln M^2/m^2$ can be 
absorbed into the effective couplings.  
The RGEs are constructed in the whole mass scales interpolating
below and beyond the decoupling scale.

In this paper, we first  generalize the method \cite{Casas:1998cf} to include wave function
renormalization in the Higgs-Yukawa model, 
and then apply it to a two real scalar model of $(\varphi, h)$
with quartic interactions. 
The mass matix of the scalar fields 
is a function of $(\varphi, h)$, and the mass eigenstates  depends on the field values. 
Thus in order to obtain the RG improved effective potential $V(\varphi, h)$, we need to use a different RGE
on a different point in the field space $(\varphi, h)$.
Furthermore,  because of the scalar mixing, the $\beta$-functions also depend on the field values.
Taking these two effects into account, we can obtain the RG improved effective potential for the 
two-scalar model. 
In this paper, we assume that the initial scale of the RGEs  
starts at a very large scale such as the Planck scale. 
Of course this is not always necessary, but such a choice helps us to understand the behavior of decoupling as we change (decrease) the renormalization scale $\mu\leq \mu_0^{}$.

The paper is organized as follows. In Section \ref{sec:review}, we   review 
and generalize the decoupling method \cite{Casas:1998cf}
in the Higgs-Yukawa model, including the wave function renormalization. 
The wave function renormalization is given by a diagram containing
 both of the scalar and the fermion fields in the loop, and their mass scales 
are generically different.  The decoupling method   determines
the decoupling scale  of each term in the effective action so that there are no threshold corrections
to the effective coupling. 
In Section \ref{sec:two scalar}, we study a two real scalar model and explicitly calculate
the RG improved effective potential. 
Section \ref{sec:summary} is devoted to summary and discussions.

\section{Higgs-Yukawa model}\label{sec:review}
In this section, we review the decoupling method of the RG improved effective potential
 proposed by Casas, Clemente and Quir\'os \cite{Casas:1998cf}\footnote{
 As we mentioned in Introduction, the idea of using the decoupling theorem was first presented in \cite{Bando:1992wy}.
 Based on the idea, 
 the authors of \cite{Casas:1998cf} made it clearer how to construct the RGEs which 
 can take into account the threshold effects by carefully introducing the decoupling scales. 
 } with a slight generalization to include wave function renormalizations.
Let us assume that we have calculated the one-loop effective potential as in Eq.(\ref{eq:general one-loop potential}).
The method adopted by the authors \cite{Casas:1998cf} is to replace  the ordinary effective
potential Eq.(\ref{eq:general one-loop potential}) by a new one with step functions: 
\aln{V'=V^{(0)}+\sum_i V^{(1)}_i\theta (V^{(1)}_i) = V^{(0)}+\sum_i V^{(1)}_i \theta (\tilde{\mu}_i^{}-M_i^{}(\{\phi_a^{}\})),
\label{effectivepotential-theta}
}
where $\theta(V_{i}^{(1)})$ is a step function which is defined to take 1 (and 0) 
at higher (lower) energy scale of $\tilde\mu$. 
When the scale $\tilde{\mu}$ is lower than the largest mass in a one-loop diagram, 
which we denote $M_i(\{ \phi_a\})$, 
the one-loop correction is set zero by the step function; 
thus the decoupling of heavy particles can be systematically taken into account. 
The effective potential is invariant under the RGE if an appropriate wave function normalization
is performed. In solving the RGEs, as we see later, 
the key identity $V \delta(V)=0$ assures absence of  further threshold corrections in the low energy
effective potential. 

In this paper we generalize to consider an effective action in order to study the 
wave function renormalization as well as renormalization of coupling constants.
Suppose we have calculated the effective action; 
\be \Gamma=S^{(0)}+\sum_{\{i\}} S^{(1)}_{\{i\}},
\label{eq:general one-loop action}
\e
where $S^{(0)}$ and $S^{(1)}_{\{i\}}$ are a tree level and one-loop effective actions respectively.
The Feynman diagrams generating the one-loop effective action generally contain various 
different particles in  loops, and
$\{ i\} $ denotes a set of particles that are contained in the loop. 
Applying the method \cite{Casas:1998cf}, we introduce step functions to incorporate the effect of 
decouplings;
\be 
 \Gamma'=S^{(0)}+\sum_i S_{\{i\}}^{(1)}  \theta(S_{\{i\}}^{(1)}) .
\label{eq:generalization}
\e
We have put a prime on the effective action to distinguish it from Eq.(\ref{eq:general one-loop action}). 
Generically $S_{\{i\}}^{(1)}$ contain multiple particles with different mass scales and thus
we cannot rewrite the step functions as  $\theta (\tilde{\mu}_i^{}-M_i^{}(\{\phi_a^{}\}))$.
One may choose the heaviest mass  $M_i^{}(\{\phi_a^{}\})$ in the loop diagram
as the renormalization scale $\tilde\mu$,
but  such a choice cannot  absorb the threshold corrections in the effective coupling constants 
 unless we expand loop integrals with respect to $1/M_i.$ 
For example, if two particles with masses, $M$ and $m$, exchange in a loop diagram,
we will have a Feynman parameter integral such as
\begin{equation}
\int_0^1 dz \log [(z M^2+ (1-z) m^2)/\mu^2 ] .
\label{Feynman}
\end{equation}
If we expanded it with respect to $m^2/M^2$, 
we would have a simple logarithmic factor $\log(M^2/\mu^2)$ with a single mass scale of the heavy field.
But it would also generate  diverging Feynman parameter integrals,
\begin{equation}
\int_0^1 dz \left( \frac{(1-z)m^2}{z M^2} \right)^n .
\end{equation}
Thus such an expansion is not justified. 
Since the integral itself (\ref{Feynman}) is convergent, 
we  treat the loop integral directly without expanding it with respect to $1/M$.
 We will see how to do this explicitly in the Higgs-Yukawa model below.

We now impose the condition that the effective action is invariant
under the following RGE; 
\aln{ &{\cal{D}}\Gamma'\equiv \left(\mu\frac{\partial}{\partial \mu}+\beta_a^{}\frac{\partial}{\partial\lambda_a^{}}-\gamma_i^{B}\phi_i^{}\frac{\partial}{\partial\phi_i^{}}-\gamma^F_i\psi\frac{\partial}{\partial\psi}-\gamma^F_i\overline{\psi}\frac{\partial}{\partial\overline{\psi}}\right)\Gamma'=0 .
\label{eq: RGE of effective action}
}
By solving the RGE (\ref{eq: RGE of effective action}), 
we can read the one-loop beta and gamma functions with the decoupling effects of 
massive particles taken into account. 
In deriving the RGE, due to the property of  $S_{\{i\}}^{(1)} \delta(S_{\{i\}}^{(1)})=0$, 
derivatives acting on the step functions vanish;
\begin{equation}
{\cal D} [S_{\{i\}}^{(1)} \theta(S_{\{i\}}^{(1)})] 
=[{\cal D} S_{\{i\}}^{(1)}] \theta(S_{\{i\}}^{(1)}) + S_{\{i\}}^{(1)} [{\cal D}\theta(S_{\{i\}}^{(1)})] 
= [{\cal D} S_{\{i\}}^{(1)}] \theta(S_{\{i\}}^{(1)}).
\end{equation} 
The absence of the $\delta$-function terms in the RGE Eq.(\ref{eq: RGE of effective action}) 
indicates that further threshold corrections are not generated in solving the equations.

Because of the step functions, 
the beta and gamma functions jump  at the decoupling scales where the step functions jump.
It reflects the fact that 
the threshold corrections  in the running couplings are cleverly absorbed in the coupling constants in the 
low energy effective theory.  
Such decoupling effects are usually put by hand in the mass independent scheme. 
The effective action is of course independent of the choice of the 
renormalization scale $\tilde{\mu}$, but 
a convenient one is
\be  \tilde\mu=\text{min}\{M_i^{}(\{\phi\}_a^{})\} .
\label{choice-of-mu}
\e
All particles are decoupled  below the scale and  no further radiative corrections arise.
As a result, 
the RG improved effective potential evaluated at the renormalization scale 
is given by the form of the tree level potential
where the coupling constants are replaced by the running couplings
calculated by using the beta functions obtained from Eq.(\ref{eq: RGE of effective action}). 
The physical meaning of this result is clear.
As long as we concentrate on scale lower than the mass of the lightest particle, 
all the effects of massive particles are absorbed in the effective couplings in the effective theory.

As a simple example, we will show how the RG improvement based on 
the beta functions derived from Eq.(\ref{eq: RGE of effective action}) 
determines the effective action of the Higgs Yukawa model \cite{Bando:1992wy}. 
The action is given by  
\be S_{\text{HY}}^{}=\int d^4x\left[\frac{1}{2}\partial_\mu^{}\phi\partial^{\mu}\phi-\frac{m^2}{2}\phi^2-\frac{\lambda}{4!}\phi^4-\Lambda+\overline{\psi}i\cancel{\partial}\psi-g\phi\overline{\psi}\psi+(\text{counter terms})
\right],
\e
where $\Lambda$ is a cosmological constant term, and we have assumed that
the fermion does not have a mass term for simplicity. 
As in \cite{Bando:1992wy} and \cite{Casas:1998cf}, we introduce
$N$ fermions in order to trace the fermion loops. 
To calculate the effective action, we 
expand each field
around  a classical configuration, $\phi=\phi_{cl}^{}+\delta \phi$, $\psi=\psi_{cl}^{}+\delta\psi$. 
Then, up to the second order of the fluctuations,
 the action becomes
\aln{ S_{\text{HY}}^{}&=S_{cl}^{}+\int d^4x\left[-\frac{1}{2}\delta\phi(\Box+M_{\phi}^{}(\phi_{cl}^{}))\delta \phi+\overline{\delta \psi}(i\cancel{\partial}-M_\psi^{}(\phi_{cl}^{}))\delta \psi-g\delta \phi\overline{\psi_{cl}^{}}\delta \psi-g\delta \phi\overline{\delta \psi}\psi_{cl}^{}+\cdots \right]
\nn
&\equiv S_{cl}^{}+\delta S_0^{}+\delta S_{int}^{}+(\text{higher order terms}),
}
where $S_{cl}^{}$ is the classical action, and
both of $\delta S_0^{} $ and $\delta S_{int}^{} $ denote  terms quadratic in the fluctuations.
Particularly, $\delta S_0$ represents terms proportional to $(\delta \phi)^2$ or  $\overline{\delta \psi} \delta \psi$
while $\delta S_{int}$ contains both fluctuations of the boson and the fermion, and thus 
induces their mixing.
Then, by treating  $\delta S_{\text{int}}$  as an interaction term, 
we get the one-loop effective action $\Gamma$ as
\aln{ 
& \exp(i\Gamma[\phi_{cl}^{},\psi_{cl}^{},\overline{\psi}_{cl}^{}])
=\exp(iS_{cl}^{})\int {\cal{D}}\delta \phi\int {\cal{D}} \delta\psi\int {\cal{D}} \overline{\delta\psi}\exp(i\delta S_0^{}+i\delta S_{int}^{})
\nn
&=\exp\left(iS_{cl}^{}+i\Gamma_{1\text{loop}}^B[\phi_{cl}]\right)\left(1+\frac{\int {\cal{D}}\delta \phi\int {\cal{D}} \delta\psi\int {\cal{D}} \overline{\delta\psi}e^{i\delta S_0^{}}(i\delta S_{int}^{}+\frac{i^2}{2}\delta S_{int}^{}\delta S_{int}^{}+\cdots)}{Z_0^{}[\phi_{cl}^{}]}\right),
}
where we defined the  one-loop effective action $\Gamma_{1\text{loop}}^B[\varphi_{cl}]$ for
a bosonic background,  i.e. $\psi_{cl}=\overline{\psi_{cl}}=0$, by  
\be Z_0^{}[\phi_{cl}^{}]\equiv \int {\cal{D}}\delta \phi\int {\cal{D}} \delta\psi\int {\cal{D}} \overline{\delta\psi}\exp\left(i\delta S_0^{}\right)\equiv \exp\left(i \Gamma_{1\text{loop}}^B[\phi_{cl}]\right).
\e
  The total one-loop effective action is then given by
\aln{
\Gamma[\phi_{cl}^{},\psi_{cl}^{},\overline{\psi}_{cl}^{}]&=S_{cl}^{}+\Gamma_{1\text{loop}}^B[\phi_{cl}]-i\ln \left(1+\frac{\int {\cal{D}}\delta \phi\int {\cal{D}} \delta\psi\int {\cal{D}} \overline{\delta\psi}e^{i\delta S_0^{}}(
\frac{i^2}{2}\delta S_{int}^{}\delta S_{int}^{}+\cdots)}{Z_0^{}[\phi_{cl}^{}]}\right).
\label{eq: one loop Higgs Yukawa}
}
In the following, we simply denote $(\phi_{cl}^{},\psi_{cl}^{})$ as $(\phi,\psi)$. After straightforward calculations, we obtain the following one-loop effective action in the $\overline{\text{MS}}$ scheme:
\aln{S^{(1)}=\int d^4x
\bigg[ &\frac{Ng^2}{16\pi^2}\phi\ln\left(\frac{M_\psi^{}(\phi)^2}{\mu^2}\right)\Box\phi+\frac{g^2}{16\pi^2}\overline{\psi}\left[G(\mu,\phi)i\cancel{\partial}-\tilde{G}(\mu,\phi)M_\psi^{}(\phi)\right]\psi\nonumber
\\
&-\frac{M_\phi^4(\phi)}{64\pi^2}\left(\ln\frac{M_\phi^2(\phi)}{\tilde{\mu}^2 }\right)+\frac{N M_\psi^4(\phi)}{16\pi^2}\left(\ln\frac{M_\psi^2(\phi)}{\tilde{\mu}^2}\right)
\bigg] .
\label{eq: one-loop effective action}
}
Here we defined
\aln{ &\tilde{\mu}^2=\mu^2e^{3/2},\ M_\phi^{2}(\phi)=m^2+\frac{\lambda}{2}\phi^2,\ \ 
 M_\psi(\phi)=g\phi,
\label{eq: Gt-integral}
}
and 
\begin{eqnarray}
G(\mu,\phi)&=&\int_0^1 dz z\ln\left(\frac{zM_\phi^2(\phi)+(1-z)M_\psi^{2}(\phi)}{\mu^2}\right), 
\nonumber  \\
  \tilde{G}(\mu,\phi)&=& \int_0^1 dz \ln\left(\frac{zM_\phi^2(\phi)+(1-z)M_\psi^{2}(\phi)}{\mu^2}\right).
\label{eq: G-integral}
\end{eqnarray}
In the calculation we have dropped higher order terms of the derivative expansion. 
See Appendix for more details of the calculations.

Following the general prescription of Eq.(\ref{eq:generalization}), 
we multiply each term of the effective action by the corresponding step function;
\aln{ 
\int d^4x
\bigg[ &\frac{Ng^2}{16\pi^2}\phi\ln\left(\frac{M_\psi^{}(\phi)^2}{\mu^2}\right)\Box\phi \ \theta_F^{}
+\frac{g^2}{16\pi^2}\overline{\psi}\left[G(\mu,\phi)i\cancel{\partial}\ \theta_G^{}-\tilde{G}(\mu,\phi)M_\psi^{}(\phi)
\  \theta_{\tilde{G}}^{}\right]\psi
\nn
&-\frac{M_\phi^4(\phi)}{64\pi^2}\left(\ln\frac{M_\phi^2(\phi)}{\tilde{\mu}^2 }\right) \ \theta_B^{}
+\frac{N M_\psi^4(\phi)}{16\pi^2}\left(\ln\frac{M_\psi^2(\phi)}{\tilde{\mu}^2}\right) \ \theta_F^{}\bigg],  
\label{eq: HY effective step}
}
where the step functions are given by
\aln{
&\theta_B^{}\equiv \theta(\tilde{\mu}-M_\phi^{}(\phi)), \nn 
& \theta_F^{}\equiv\theta(\tilde{\mu}-|M_\psi^{}(\phi)|),
\nn
& \theta_G^{}\equiv \theta(G(\mu,\phi))=\theta(\mu-\mu_G^{}(\phi)), \nn 
& \theta_{\tilde{G}}^{}\equiv \theta(\tilde{G}(\mu,\phi))=\theta(\mu-\mu_{\tilde{G}}^{}(\phi)). 
\label{stepfunctions4}
}
Without loss of generality, we can assume $M_\psi^{}(\phi)>0$ in the following.
The integrations over the Feynman parameter $z$ in $G$ and $\tilde{G}$
can be explicitly performed, and we can translate $\theta_G$ and $\theta_{\tilde{G}}$
into step functions with 
 the decoupling scales,  $\mu_G^{}(\phi)$ and $\mu_{\tilde{G}}^{}(\phi)$;
\aln{&\mu_G^{}(\phi)=M_\phi^{}(\phi)\times \exp\left(-\frac{1-3A^2}{4(1-A^2)}+\frac{A^4\ln A}{(1-A^2)^2}\right),
\\
&\mu_{\tilde{G}}^{}(\phi)=M_\phi^{}(\phi)\times \exp\left(-\frac{1}{2}-\frac{A^2\ln A}{1-A^2}\right)   ,
}
where $A\equiv  M_\psi^{}(\phi)/M_\phi^{}(\phi)$.
\begin{figure}[t!]
\begin{center}
\includegraphics[width=9cm]{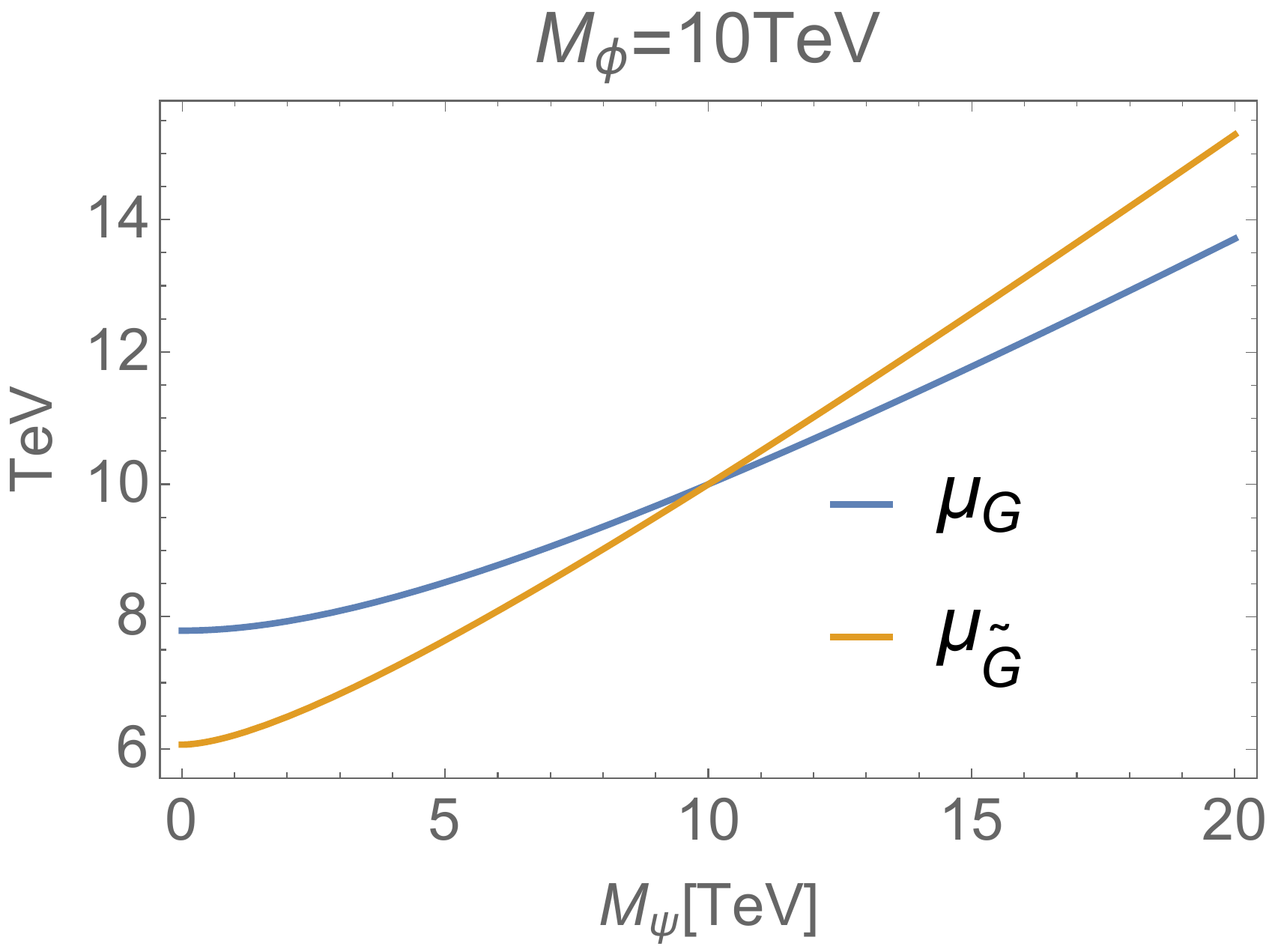}
\caption{The decoupling scales $\mu_G^{}$ and $\mu_{\tilde{G}}^{}$ as functions of $M_\psi^{}(\phi)$. Here, $M_\phi^{}(\phi)$ is fixed to be $10$TeV.  The decoupling scales are in between the mass scales of a heavy and a light particle
and never coincide with the heavier mass scale.
}
\label{fig:decoupling scale}
\end{center}
\end{figure}
In Fig.\ref{fig:decoupling scale}, we plot these decoupling scales 
as  functions of $M_\psi^{}(\phi)$ where $M_\phi^{}(\phi)$ is fixed to be $10$ TeV. 
They increase as $M_\psi$ increases and coincide with $M_\psi$ when 
it is equal to $M_\phi=10 \ \text{TeV}$. For $M_\psi > M_\phi$ the decoupling scales satisfy
$M_\psi >\mu_{\tilde{G}}>\mu_G >  M_\phi$, and $M_\psi < \mu_{\tilde{G}} < \mu_G <  M_\phi$
for $M_\psi < M_\phi$. 
Note that they  are well-behaved in the potentially dangerous limits $A\rightarrow 0, \infty$.
In the $A \rightarrow \infty$ limit,
\begin{eqnarray}
&\mu_G^{}(\phi) \rightarrow  M_\psi^{}(\phi) e^{-3/4} , \ \ 
\mu_{\tilde{G}}^{}(\phi) \rightarrow M_\psi^{}(\phi) e^{-1/2}
\end{eqnarray}
and in the $A \rightarrow 0$ limit,
\begin{eqnarray}
&\mu_G^{}(\phi) \rightarrow  M_\phi^{}(\phi) e^{-1/4} , \ \ 
\mu_{\tilde{G}}^{}(\phi) \rightarrow M_\phi^{}(\phi) e^{-1/2} .
\end{eqnarray}
They do not  coincide exactly with $M_\phi^{}(\phi)$ or $M_\psi^{}(\phi)$
 in the $M_\psi^{}(\phi)\rightarrow 0 $ or $\infty$ limit because of the Feynman 
 parameter integral in Eqs.(\ref{eq: Gt-integral})(\ref{eq: G-integral}): the decoupling scale
is a bit smaller than the mass scale of a heavier particle. 
When one of them is much heavier than the lighter one, 
the decoupling scale is affected by the presence of the light particle in the loop. 
Nonetheless the effect is finite.
The finiteness is physically natural, but 
it is not trivial to prove that it is maintained in higher loop corrections. 
In \cite{Casas:1998cf}, it was explicitly shown that the decoupling scales appearing in the effective potential are well-behaved up to two loop order. 
The fact that a heavy particle is not decoupled at the mass but at a lower energy scale 
will be important in probing physics beyond the SM.

We can now read the one-loop $\beta$ and $\gamma$ functions from the RGE 
Eq.(\ref{eq: RGE of effective action});
\aln{ &\beta_\lambda^{}=\frac{d\lambda}{d\log\mu}
=\frac{1}{16\pi^2}\left(3\lambda^2\theta_B^{}+
(8N\lambda g^2 -48Ng^4) \theta_F^{}\right)
,\label{eq:one-loop RGEs 1}
\\
&\beta_g^{}=\frac{dg}{d\log\mu}=\frac{g^3}{16\pi^2}
\left(
  \theta_G^{}+2\theta_{\tilde{G}}^{}
 +2N \theta_F^{}\right)
 ,\label{eq:one-loop RGEs 2}
\\
&\beta_{m^2}^{}=\frac{dm^2}{d\log\mu}=\frac{m^2}{16\pi^2}(\lambda\theta_B^{}+4Ng^2\theta_F^{})
,\label{eq:one-loop RGEs 3}
\\
&\gamma^{B}=\frac{2Ng^2}{16\pi^2}\theta_F^{},
\\
&\gamma^{F}=\frac{g^2}{32\pi^2}\theta_G^{}. 
\label{eq:one-loop RGEs 4}
}
\begin{figure}
\begin{minipage}{0.5\hsize}
\begin{center}
\includegraphics[width=9cm]{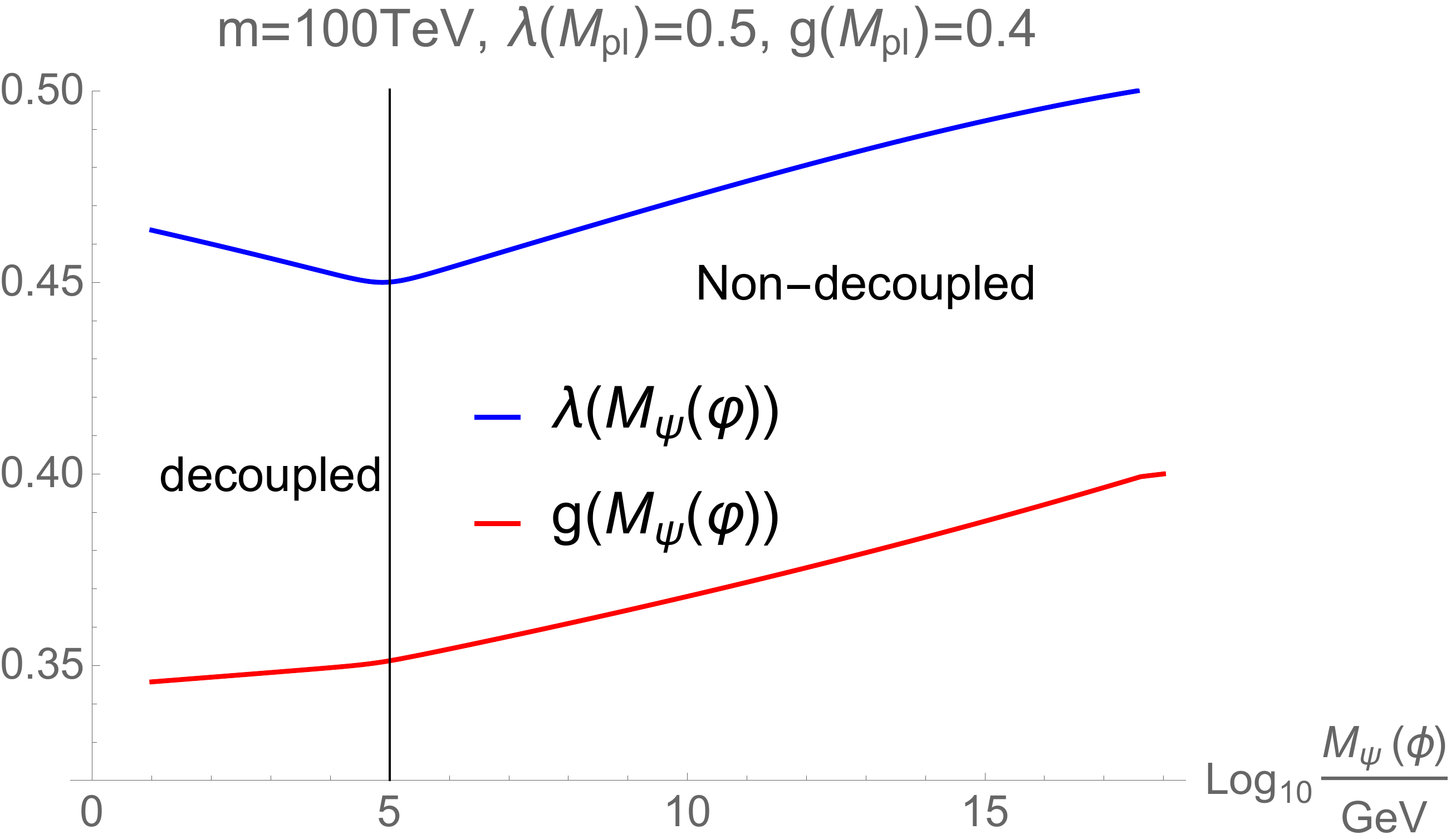}
\end{center}
\end{minipage}
\begin{minipage}{0.5\hsize}
\begin{center}
\includegraphics[width=8cm]{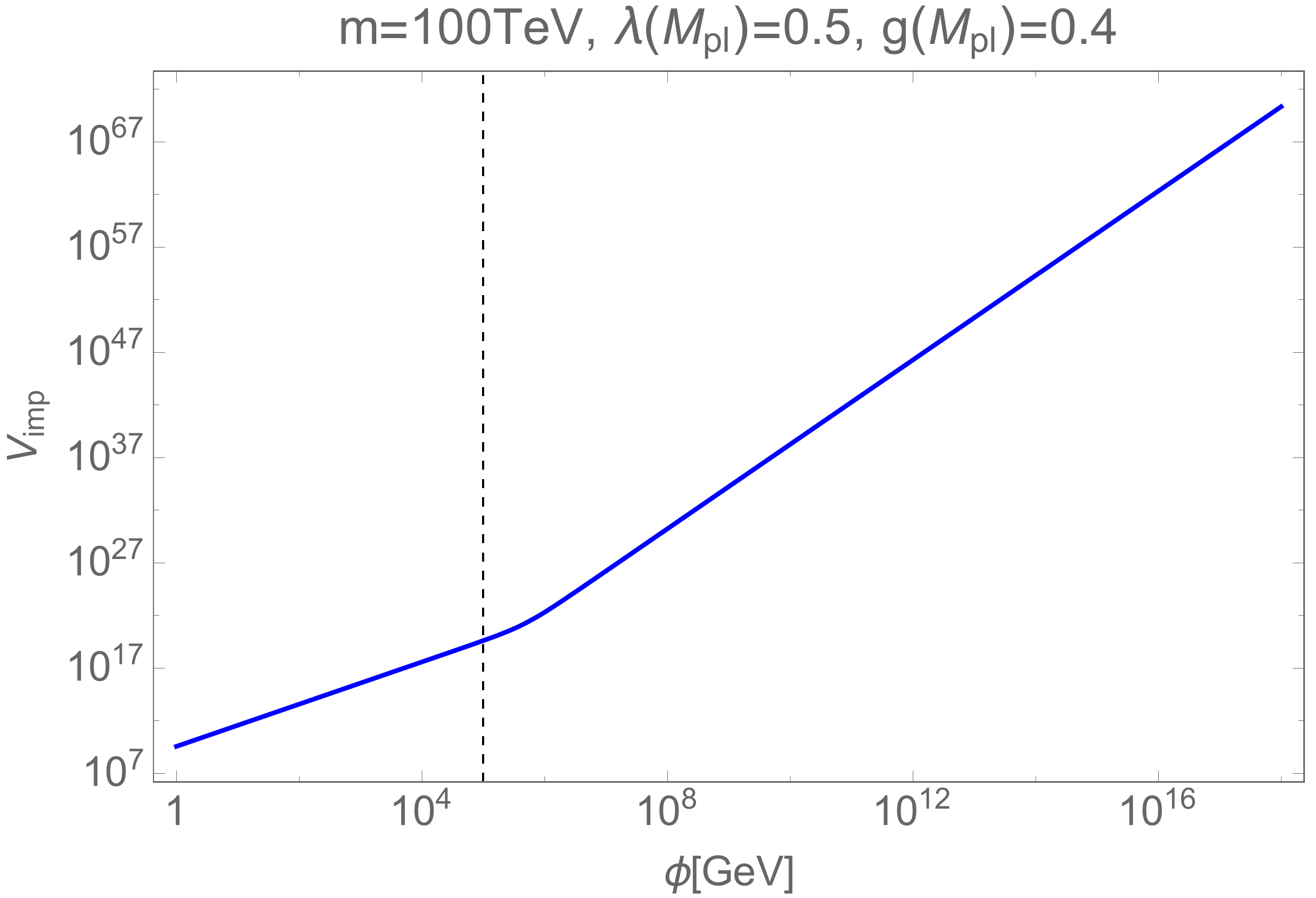}
\end{center}
\end{minipage}
\caption{Left: The running couplings at $\tilde{\mu}=M_\psi^{}(\phi)$ are evaluated by using Eqs.(\ref{eq:one-loop RGEs 1})-(\ref{eq:one-loop RGEs 4}). Here, we set $\lambda(M_{pl}^{})=0.5$, $g(M_{pl}^{})=0.4$ and $m(M_{pl}^{})=100$TeV 
as the initial values. Right: The corresponding improved one-loop effective potential. 
} 
\label{fig:HY}
\end{figure}
\noindent By solving the running couplings from the above RGEs, we can obtain the 
improved  effective potential.
 For example, in the region $g\phi\gtrsim m$,  as far as $\lambda\sim g^2$, 
both of the boson and the fermion have similar scales of mass  
$M_\phi^{}(\phi)\sim M_\psi^{}(\phi)$. 
Then both of these fields contribute to the RGEs similarly and decouple around the same energy scale. 
Therefore, the resultant running couplings at $\tilde{\mu}= M_\phi^{}(\phi) \sim M_\psi^{}(\phi)$ 
coincide with the ordinary ones that are obtained by solving the usual RGEs without step functions. 
On the other hand, if $g\phi\lesssim m$, 
the scalar field is heavier than the fermion $M_\phi^{}(\phi)\gtrsim M_\psi^{}(\phi)$
and decouples at $M_\phi^{}(\phi)$. 
Below this scale we have only the fermionic contributions to the RGE. 
This behavior is schematically shown in the left panel of Fig.\ref{fig:HY}. 
As a result, the tree-level potential 
\be V(\phi)=\overline{\Lambda}(\mu)+\frac{\overline{m}^2(\mu)}{2}\overline{\phi}(\mu)^2+\frac{\overline{\lambda}(\mu)}{4!}\overline{\phi}(\mu)^4 \bigg|_{\tilde{\mu}=M_\psi^{}(\phi)}^{}
\e
whose running couplings are evaluated at $\tilde{\mu}=\text{min}(M_\phi^{}(\phi),M_\psi^{}(\phi))$ at each value of $\phi$
coincides with the result in \cite{Bando:1992wy}. 

As shown in Eqs.(\ref{eq:one-loop RGEs 1})-(\ref{eq:one-loop RGEs 4}), the beta and gamma functions
are discontinuous at the decoupling scales. 
Thus the couplings obey different RGEs in each region. However, it can be straightforwardly shown
that these seemingly different RGEs describe the same equation and are related by a change of variables. 
If $g\phi\leq m$ and $M_\psi^2(\phi) < M_\phi^2(\phi)$ is satisfied, the boson is decoupled at  higher
energy scale and we can expand 
 the bosonic part of the one-loop potential in Eq.(\ref{eq: one-loop effective action}) with respect to $\phi/m$.
Then various parameters in the action are modified \cite{Bando:1992wy};
\footnote{Note that, in this naive choice of the decoupling scale $\tilde{\mu}=m$, we have finite threshold corrections to the 
coupling constants, e.g. $\delta g=g^3/16\pi^2$. On the other hand, in our present method, because we define the decoupling scales by the scales where the one-loop corrections vanish, there are no threshold corrections. 
For example, for the Yukawa term, after the field redefinition $(1+g^3G/16\pi^2)^{1/2}\psi\rightarrow \psi$, 
the one-loop correction can be rewritten as 
\aln{ -\int d^4x\frac{g^3}{16\pi^2}\left(G+\tilde{G}\right)\phi\overline{\psi}\psi
= -\int d^4x\frac{g^3}{16\pi^2}\left[\frac{1}{2}\left(\ln\frac{\mu_{G}^2}{\mu^2}\right)+\left(\ln\frac{\mu_{\tilde{G}}^2}{\mu^2}\right)\right]\phi\overline{\psi}\psi, \nonumber
}
which corresponds to the modified coupling constant 
$g'=g+g^3\left(\frac{1}{2}\ln \mu_{G}^2/\mu^2+\ln \mu_{\tilde{G}}^2/\mu^2 \right)/ 16\pi^2$ 
below the decoupling scales of $\mu_G^{}$ and $\mu_{\tilde{G}}^{}$. 
We can obtain the simplified transformation of Eq.(\ref{eq: yukawa relation}) 
by expanding $g^3 \left(\frac{1}{2}\ln \mu_{G}^2/\mu^2+\ln\ \mu_{\tilde{G}}^2/\mu^2 \right)/16\pi^2$ 
as a function of $\phi/m$.
}
\aln{ & m'^2=m^2\left(1 + \frac{\lambda}{32\pi^2}\left( \frac{1}{2} +
\ln\frac{m^2}{\tilde{\mu}^2}\right)\right),
\label{eq: mass relation}
\\
&\lambda'=\lambda\left(1+\frac{3\lambda}{32\pi^2}
\left( \ln\frac{m^2}{\tilde{\mu}^2}+ \frac{3}{2}  \right) \right),
\label{eq: lambda relation}
\\
&\Lambda'=\Lambda+\frac{m^4}{64\pi^2}\left(\ln\frac{m^2}{\tilde{\mu}^2}\right),
\label{eq: Lambda relation}
\\
& 
g'=g+\frac{g^3}{16\pi^2}\left(\frac{3}{2}\ln\frac{m^2}{\tilde{\mu}^2}+1\right).
\label{eq: yukawa relation}
}
For simplicity, let us assume that the bosonic contributions to the RGEs are all decoupled at the same scale $M_\phi^{}\sim \mu_G^{}\sim \mu_{\tilde{G}}^{}$, and denote the beta functions in Eqs.(\ref{eq:one-loop RGEs 1})-(\ref{eq:one-loop RGEs 4}) 
 below $\tilde{\mu}< M_\phi$ (but above $M_\psi^{})$  as $\beta_{\lambda'}$ with a prime
 and those above $\tilde{\mu} >M_\phi > M_\psi$ as $\beta_\lambda$ without a prime.  
Then, by using Eqs.(\ref{eq: mass relation})-(\ref{eq: yukawa relation}), 
the differential operator ${\cal D}$ 
defined in Eq.(\ref{eq: RGE of effective action}) is shown \cite{Bando:1992wy}  to be identical with the new differential operator ${\cal{D}}'$;
\aln{
{\cal{D}}&=\left(\frac{\partial}{\partial t}+({\cal{D}}\lambda')\frac{\partial}{\partial\lambda'}+({\cal{D}}g')\frac{\partial}{\partial g'}+({\cal{D}}m'^2)\frac{\partial}{\partial m'^2}+({\cal{D}}\Lambda')\frac{\partial}{\partial \Lambda'}-\gamma^B\phi\frac{\partial}{\partial \phi}\right)
\nn
&=\left(\frac{\partial}{\partial t}+\beta_{\lambda'}^{}\frac{\partial}{\partial\lambda'}+\beta_{g'}^{}\frac{\partial}{\partial g'}+\beta_{m'^2}^{}\frac{\partial}{\partial m'^2}+\beta_{\Lambda'}^{}\frac{\partial}{\partial \Lambda'}-\gamma^B\phi\frac{\partial}{\partial \phi}\right)={\cal{D}}'.
}
More generally, 
when we distinguish the several decoupling scales, we have  seemingly different differential operators $({\cal{D}},{\cal{D}}',{\cal{D}}'',\cdots )$
 in each interval between the decoupling scales. But 
 they all describe the same RGE by the same reasoning above. 

The above construction of the improved action is quite general and can be applicable to any field theory with arbitrary number of mass scales. In a theory with only one scalar field, the structure of the improvement is almost the same as the above Higgs-Yukawa model: Introduce step functions in the RGEs 
and evaluate running couplings at $\tilde{\mu}=\text{min}(M_i^{}(\phi))$. This gives the RG improved effective action
including the threshold corrections.
But the construction becomes practically involved in models with multi scalar fields 
because 
(i) mass eigenvalues generally depend on several scalar fields and 
(ii) scalar mixing couplings generate the non-polynomial terms in the effective potential. 
In the following,  we consider a two real scalar model as another example.
Such a model  is phenomenologically well motivated and
it will be important to obtain the RG improved effective action in the whole region of the field values
of the two scalars.

\section{Two real scalar model}\label{sec:two scalar}
In this section, we calculate the RG improved effective potential of a two real scalar model 
based on the method presented in the previous section. 
Since there are no wave function renormalization at one-loop order, we can concentrate on the
effective potential. 
The action is given by
\be {\cal{L}}=\frac{1}{2}(\partial_\mu^{}\varphi)^2+\frac{1}{2}(\partial_\mu^{}h)^2-\frac{\lambda_\varphi^{}}{4!}\varphi^4-\frac{\kappa}{4}\varphi^2h^2-\frac{\lambda_h^{}}{4!}h^4 .
\label{eq:one loop two scalar}
\e
In the following, we denote the tree level potential as $V^{(0)}$. In the $\overline{\text{MS}}$ scheme, 
the one-loop effective potential is calculated as
\aln{ &V(\varphi,h)=V^{(0)}+V^{(1)},\nonumber
\\
&V^{(1)}(\varphi,h; \mu)=\frac{1}{64\pi^2}\left[M_+^4\left(\log\left(\frac{M_+^2}{\tilde{\mu}^2}\right)\right)\theta(\tilde{\mu}-M_+^{})
+M_-^4\left(\log\left(\frac{M_-^2}{\tilde{\mu}^2}\right)\right)\theta(\tilde{\mu}-M_-^{})
\right]
\label{V1(phih)}
}
where
\aln{ M_+^2&=\frac{1}{4} \left(h^2 \left(\lambda _h+\kappa \right)+\varphi ^2 \left(\kappa +\lambda _{\varphi
   }\right)
+\tilde{M}^2\right),
   \\ 
   M_-^2&=\frac{1}{4} \left(h^2 \left(\lambda _h+\kappa \right)+\varphi ^2 \left(\kappa +\lambda _{\varphi
   }\right)
   -\tilde{M}^2\right),
   \\
   \tilde{M}^2&=\sqrt{h^4 \left(\kappa -\lambda
   _h\right){}^2+2 h^2 \varphi ^2 \left(\kappa  \left(\lambda _h+\lambda _{\varphi
   }\right)-\lambda _h \lambda _{\varphi }+7 \kappa ^2\right)+\varphi ^4 \left(\kappa
   -\lambda _{\varphi }\right){}^2}.
}
In this model, there is no mixture of different particles in a single one-loop diagram
and the decoupling scales are simply given by the mass scale of the heavy particle.
But there is another complication. 
As mentioned at the end of the previous section, the effective potential contains a non-polynomial
term of the scalar field values through $\tilde{M}$ and we need to expand it to obtain
polynomial potentials. 
The one-loop beta functions can be read from the condition that the effective action should be
independent on the renormalization scale;
\begin{eqnarray}
 &&  \left(\mu\frac{\partial}{\partial\mu}
+\beta_{\lambda_\varphi^{}}\frac{\partial}{\partial\lambda_\varphi^{}}
+\beta_{\lambda_h^{}}\frac{\partial}{\partial\lambda_h^{}}
+\beta_{\kappa}\frac{\partial}{\partial\kappa}
-\gamma_\varphi^{}\varphi\frac{\partial}{\partial\varphi}
-\gamma_h^{}h\frac{\partial}{\partial h}\right)V(\varphi,h) \nonumber \\
&& =  
 \mu\frac{\partial}{\partial\mu}V^{(1)}(\varphi,h)+\left(\beta_{\lambda_\varphi^{}}\frac{\partial}{\partial\lambda_\varphi^{}}
+\beta_{\lambda_h^{}}\frac{\partial}{\partial\lambda_h^{}}
+\beta_{\kappa}\frac{\partial}{\partial\kappa}\right)V^{(0)}(\varphi,h) =0.
\label{eq: two scalar RGE}
\end{eqnarray}
Here we have used the fact that the gamma functions vanish at the one-loop order in the present model. 
From Eq.(\ref{V1(phih)}), 
the $\mu$-derivative of $V^{(1)}$ is given by
\aln{
 \mu\frac{\partial}{\partial\mu}V^{(1)}(\varphi,h)
=& \ -\frac{1}{2}\left(\frac{\beta_{\lambda_\varphi^{}}^{(1)}}{4!}\varphi^4+\frac{\beta_{\lambda_h^{}}^{(1)}}{4!}h^4
+\frac{\beta_{\kappa^{}}^{(1)}}{4}\varphi^2h^2
\right)\left[\theta(\tilde{\mu}-M_+^{})+\theta(\tilde{\mu}-M_-^{})\right]\nonumber
\\
& -\frac{1}{32\pi^2\cdot 8}\{h^2 \left(\lambda _h+\kappa \right)+\varphi ^2 \left(\kappa +\lambda _{\varphi
   }\right)\}\tilde{M}^2\left[\theta(\tilde{\mu}-M_+^{})-\theta(\tilde{\mu}-M_-^{})\right]
\label{mu-derivative-ofV1}
}
where $\beta_{\lambda_\varphi^{}}^{(1)}$, $\beta_{\lambda_h^{}}^{(1)}$ and $\beta_{\kappa^{}}^{(1)}$ are 
 ordinary one-loop beta functions,
\aln{\beta_{\lambda_\varphi^{}}^{(1)}&=\frac{3}{16\pi^2}\left(\lambda_\varphi^2+\kappa^2\right),\nonumber
\\
\beta_{\lambda_h^{}}^{(1)}&=\frac{3}{16\pi^2}\left(\lambda_h^2+\kappa^2\right),
\label{eq:ordinary scalar RGEs}
\\
\beta_{\kappa}^{(1)}&=\frac{\kappa}{16\pi^2}\left(\lambda_h+4\kappa+\lambda_\varphi\right).
\nonumber
}
If two mass scales $M_\pm^2$ are equal, the second term in Eq.(\ref{mu-derivative-ofV1}) vanishes
and we get the ordinary beta functions $\beta_\star=\beta_\star^{(1)}$ for $\star=\lambda_\varphi, \lambda_h, \kappa$.  
But if there is a hierarchy between these two mass scales,
we need to take the effect of $M_+^{} \neq M_-^{}$ in the calculation of the RG improved effective action.
By expanding the non-polynomial term $\tilde{M}$ and 
inserting Eq.(\ref{mu-derivative-ofV1}) into Eq.(\ref{eq: two scalar RGE}), we can obtain 
the beta functions $\beta_\star^{}$ in which the decoupling effects are automatically
taken into account. 

\begin{figure}
\begin{center}
\includegraphics[width=9cm]{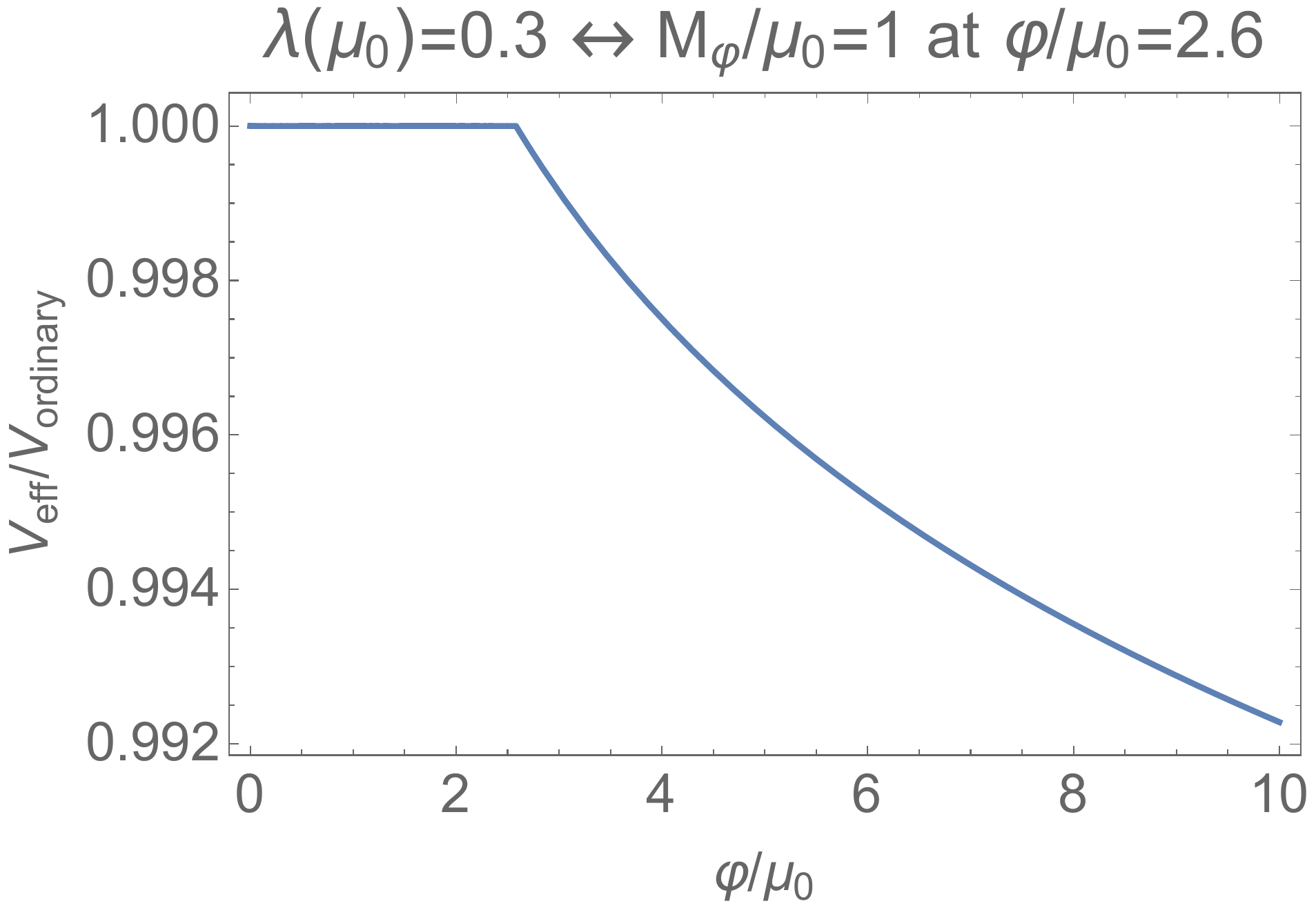}
\caption{The ratio of  $V(\varphi)/V_{\text{ordinary}}^{}(\varphi)$ is plotted as a function of $\varphi/\mu_0$
in the case of $\kappa=0$.  
As the initial boundary condition, we set $\lambda_\varphi^{}(\mu_0^{})=0.3$.
This corresponds to the decoupling scale $\varphi_{\text{dec}}^{}/\mu_0^{}\simeq 2.6$. 
We can see that $V(\varphi)$  coincides with $V_{\text{ordinary}}^{}(\varphi)$ 
for $\varphi < \varphi_0$, but it becomes different when $\varphi > \varphi_0$ 
and the particle is decoupled from the RGE at $\varphi_0.$
} 
\label{fig:comparison of potentials}
\end{center}
\end{figure}

Before considering  a generic case of $\kappa\neq0$, let us first study the $\kappa=0$ case. 
In this case,  there is no scalar mixing and the masses are simply given by
\begin{eqnarray}
M_+ = \sqrt{\frac{\lambda_\varphi^{}}{2}h^2}, \ \  
M_- = \sqrt{\frac{\lambda_h^{}}{2}h^2}, \  \ 
\tilde{M} =\lambda_h^{}h^2-\lambda _{\varphi}^{}\varphi^2 .
\end{eqnarray}
Thus we have
\begin{equation}
 \mu\frac{\partial}{\partial\mu}V^{(1)}(\varphi,h)
= -\frac{\beta_{\lambda_\varphi^{}}^{(1)}}{4!}\varphi^4
\theta\left(\tilde{\mu}-\sqrt{\frac{\lambda_\varphi^{}}{2}}\varphi^2\right)
-\frac{\beta_{\lambda_h^{}}^{(1)}}{4!}h^4\theta\left(\tilde{\mu}-\sqrt{\frac{\lambda_h^{}}{2}h^2}\right).
\end{equation}
Then, by inserting it into Eq.(\ref{eq: two scalar RGE}),
 the beta functions are given by
\begin{equation}
\beta_{\lambda_\varphi^{}}^{}= \frac{3  \lambda_\varphi^2}{16\pi^2}
\theta\left(\tilde{\mu}-\sqrt{\frac{\lambda_\varphi^{}}{2}\varphi^2}\right),
\ \ \beta_{\lambda_h^{}}^{}=\frac{3  \lambda_h^2}{16\pi^2}
\theta\left(\tilde{\mu}-\sqrt{\frac{\lambda_h^{}}{2}h^2}\right).
\label{eq: kappa zero case}
\end{equation}
If we choose the renormalization scale as
\be \tilde{\mu}=\text{min}\left(\sqrt{\frac{\lambda_h\varphi^{}}{2}\varphi^2},\sqrt{\frac{\lambda_h^{}}{2}h^2}\right)\equiv \text{min}\left(M_\varphi^{},M_h^{}\right),
\e
we obtain the RG improved effective potential;
\be V(\varphi,h)=\frac{\lambda_\varphi^{}\left(\tilde{\mu}=M_\varphi^{}\right)}{4!}\varphi^4+\frac{\lambda_h^{}\left(\tilde{\mu}=M_h^{}\right)}{4!}h^4. 
\label{eq: kappa0 improvement}
\e
The running couplings are evaluated by using Eq.(\ref{eq: kappa zero case}). 
This result agrees with the ordinary RG improved potential,
but the RGE itself is different.
 In the present case, the coupling constants remain at their initial values 
 in the region $\mu_0^{}\leq M_{\varphi(h)}^{}$ because of the step functions in Eq.(\ref{eq: kappa zero case}). 
 In Fig.\ref{fig:comparison of potentials},  we plot $V(\varphi,0)/V_{\text{ordinary}}^{}(\varphi)$
 where $V_{\text{ordinary}}^{}(\varphi)$ is the ordinary improved potential without a step function in $\beta_\lambda^{}$. 
In this figure, we use $\lambda_\varphi^{}(\mu_0^{})=0.3$ as the boundary condition of $\lambda_\varphi^{}$.
Then the step function becomes $\theta(\tilde{\mu}-\sqrt{0.3/2} \varphi)$ 
and the decoupling scale $\tilde{\mu}=\mu_0$ 
is given by $\varphi_{\text{dec}}^{}/\mu_0^{}\simeq 2.6$. 
Of course these two effective potentials 
 coincide for $\varphi<\varphi_{\text{dec}}^{}$. 
 Thus, as long as we take a very high scale $\Lambda$ such as the Planck scale as an initial scale of RGEs, the present method correctly reproduces the usual improvement in the low energy region $M_{\varphi}^{} (\varphi)\lesssim \Lambda\sim M_{pl}^{}$. 
\\

We now consider a generic case of $\kappa \neq 0$. 
We first expand $\tilde{M}^2$ in Eq.(\ref{eq: two scalar RGE}) as a function of
 $\varphi$ and $h$ in order to obtain the beta functions. 
It is  similar to the $(\phi/m)^2$ expansion in the Higgs-Yukawa model 
which is valid when the condition 
$\phi^2 \ll m^2$ is satisfied. 
In the present case, expansions of $\tilde{M}^2$  should be different in  different regions
of $(\varphi, h)$.
 For example, in a region satisfying $\varphi\gg h$, 
 we  can expand $\tilde{M}^2$  with respect to $(h/\varphi)^2$. 
 More generally, if we are interested in the effective potential around 
the following classical configuration ($r \neq 0$), 
\be 
 \varphi_{cl}^2 = r^2 \cos\theta,\ h_{cl}^2=r^2 \sin\theta,
\label{classical-config}
\e
we can separate field variables into large and small components 
by defining new variables as
\be \left(\begin{array}{c}X^2 \\ Y^2
\end{array}\right)=\left(\begin{array}{cc}\cos \theta  &\sin\theta
\\
-\sin\theta & \cos \theta 
\end{array}\right)\left(\begin{array}{c}\varphi^2 \\ h^2
\end{array}\right) 
\label{X2Y2}
\e
where $0 \le \theta \le \pi/2.$ Without loss of generality, we can assume that both of the field values
$(\varphi, h)$ or $(X,Y)$ are positive. 
For the classical configuration Eq.(\ref{classical-config}), 
$X_{cl}^2=r^2$ and $Y_{cl}^2=0$ are satisfied. 
Thus, around the classical solution, $X^2$ is the large component and we can safely expand
$\tilde{M}$ with respect to $Y^2/X^2$. 
Substituting 
$\varphi^2=\varphi^2(X^2,Y^2)$ and $h^2=h^2(X^2,Y^2)$ into $\tilde{M}^2$ and expanding it 
with respect to $Y^2/X^2$, 
we obtain 
\begin{eqnarray}
\tilde{M}^2 
 &\approx &
 \left[ X^2 \left( (7 \kappa^2 -\lambda_\varphi \lambda_h + \kappa (\lambda_\varphi +\lambda_h) )\sin 2 \theta+
 (\kappa-\lambda_\varphi)^2 \cos^2 \theta + (\kappa-\lambda_h)^2 \sin^2 \theta
 \right)
\right. \nonumber \\
&& \left. + Y^2 \left( (7 \kappa^2 -\lambda_\varphi \lambda_h + \kappa (\lambda_\varphi +\lambda_h) )\cos 2 \theta
+ (\lambda_\varphi -\lambda_h)(\lambda_\varphi +\lambda_h-2\kappa) \cos \theta \sin \theta \right)  \right]
\nonumber \\
&& \times \frac{1}{\sqrt{(\kappa-\lambda_\varphi)^2 \cos^2 \theta + (\kappa-\lambda_h)^2 \sin^2 \theta
+ (7 \kappa^2 -\lambda_\varphi \lambda_h + \kappa (\lambda_\varphi +\lambda_h) )\sin 2 \theta}}
\nonumber \\
\end{eqnarray}
up to ${\cal O}(Y^4/X^2)$. Using Eq.(\ref{X2Y2}) and inserting it
back to Eq.(\ref{mu-derivative-ofV1}), 
we obtain the beta functions around the classical solution Eq.(\ref{classical-config});
\begin{eqnarray}
 \beta_{\lambda_h}
&=& \frac{1}{2}\beta_{\lambda_h^{}}^{(1)}[\theta(\tilde{\mu}-M_+^{})+\theta(\tilde{\mu}-M_-^{})]
\nonumber  \\
&& +\frac{3(\kappa+\lambda_h)}{32\pi^2} 
\left( 
\frac{ (7 \kappa^2 -\lambda_\varphi \lambda_h + \kappa (\lambda_\varphi +\lambda_h) ) \cos \theta
+ (\kappa -\lambda_h)^2 \sin \theta }
{\sqrt{(\kappa-\lambda_\varphi)^2 \cos^2 \theta + (\kappa-\lambda_h)^2 \sin^2 \theta
+ (7 \kappa^2 -\lambda_\varphi \lambda_h + \kappa (\lambda_\varphi +\lambda_h) )\sin 2 \theta}} \right)
\nonumber
   \\
 &&  \times [\theta(\tilde{\mu}-M_+^{})-\theta(\tilde{\mu}-M_-^{})],
 \label{eq: scalar beta 1}
   \\
 \beta_{\lambda_\varphi}
&=& \frac{1}{2}\beta_{\lambda_\varphi}^{(1)}[\theta(\tilde{\mu}-M_+^{})+\theta(\tilde{\mu}-M_-^{})]
\nonumber  \\
&& +\frac{3(\kappa+\lambda_\varphi)}{32\pi^2} 
\left( 
\frac{ (7 \kappa^2 -\lambda_\varphi \lambda_h + \kappa (\lambda_\varphi +\lambda_h) ) \sin \theta
+ (\kappa -\lambda_\varphi)^2 \cos \theta }
{\sqrt{(\kappa-\lambda_\varphi)^2 \cos^2 \theta + (\kappa-\lambda_h)^2 \sin^2 \theta
+ (7 \kappa^2 -\lambda_\varphi \lambda_h + \kappa (\lambda_\varphi +\lambda_h) )\sin 2 \theta}} 
\right)
\nonumber
   \\
 &&  \times [\theta(\tilde{\mu}-M_+^{})-\theta(\tilde{\mu}-M_-^{})],
 \label{eq: scalar beta 2}
\\ 
\beta_{\kappa^{}}^{}&=& \frac{1}{2}\beta_{\kappa^{}}^{(1)}[\theta(\tilde{\mu}-M_+^{})+\theta(\tilde{\mu}-M_-^{})]\nonumber
\\
&&+\frac{\kappa}{32\pi^2}
\bigg(
\frac{(4 \kappa^2 +\lambda_\varphi (\lambda_\varphi- \lambda_h) +\kappa (3 \lambda_\varphi + \lambda_h)) \cos \theta
+ (4 \kappa^2 +\lambda_h (\lambda_h- \lambda_\varphi) +\kappa ( \lambda_\varphi + 3 \lambda_h)) \sin \theta}
 {\sqrt{(\kappa-\lambda_\varphi)^2 \cos^2 \theta + (\kappa-\lambda_h)^2 \sin^2 \theta
+ (7 \kappa^2 -\lambda_\varphi \lambda_h + \kappa (\lambda_\varphi +\lambda_h) )\sin 2 \theta}}   
\bigg)
   \nonumber \\
 && \times [\theta(\tilde{\mu}-M_+^{})-\theta(\tilde{\mu}-M_-^{})].
 \label{eq: scalar beta 3}
\end{eqnarray}
These beta functions are different from the ordinary ones Eq.(\ref{eq:ordinary scalar RGEs})
when $M_+ \neq M_+$ because the decoupled component with mass $M_+$ is different 
for a different value of $\theta$. The second terms in the beta functions represent
the effects of the lighter particles with mass $M_-$, and contribute to the RGEs during
the scale $M_- < \tilde{\mu} < M_+$.
Note that the coefficients of the second terms depend on $\theta = \text{arctan}(h^2/\varphi^2)$.
\begin{figure}
\begin{center}
\includegraphics[width=9cm]{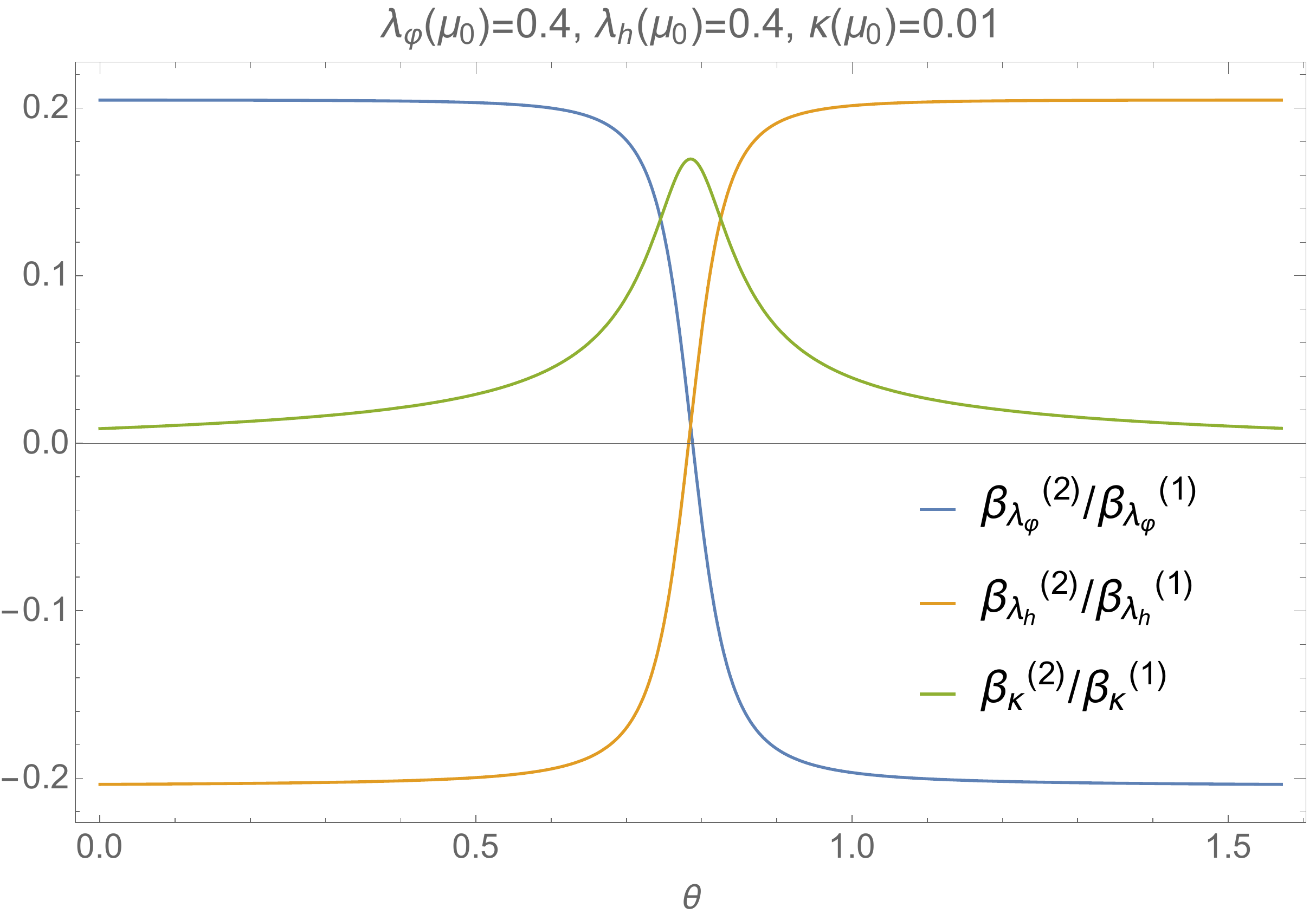}
\caption{The  second terms in the beta functions Eqs.(\ref{eq: scalar beta 1})-(\ref{eq: scalar beta 3}) are plotted 
as functions of $\theta$. Here we normalize them by their first terms respectively.  
} 
\label{fig:beta_theta}
\end{center}
\end{figure}

In Fig.\ref{fig:beta_theta}, we show the second terms in the beta functions as  functions of $\theta$, 
 normalized by the first terms, i.e. the ordinary beta functions Eq.(\ref{eq:ordinary scalar RGEs}).   
As we saw in Fig.\ref{fig:beta_theta}, 
the effect can be sizable if there is a large hierarchy between the two mass scales. 
For example, suppose
 $|\kappa|\ll 1$ and $\theta\sim 0$ for simplicity.  Along the direction $\theta=0$, we have
\aln{ M_+^{}=\frac{\sqrt{\lambda_\varphi^{}}}{2}\varphi,\ \ \ M_-^{}=\frac{\sqrt{\kappa}}{2}\varphi .
 } 
If $\lambda_\varphi^{}={\cal{O}}(0.1)$, we have
 $\ln (M_+^{}/M_-^{})\sim (\ln \kappa^{-1})/2$. 
Hence, if we take the Planck scale $M_{pl}^{}$ as the initial scale of the RGE 
and $M_-^{}$ at the EW scale $\sim 100$GeV,  
$
\beta_{\lambda}^{(2)} \ln(M_+^{}/M_-^{})
$
gives a typical order of corrections to the low energy effective couplings below the EW scale
due the step functions in the beta functions. 
Here $\beta_{\lambda}^{(2)}$ denotes the coefficient of $[\theta(\tilde{\mu}-M_+^{})-\theta(\tilde{\mu}-M_-^{})]$ in the 
beta function Eq.(\ref{eq: scalar beta 2}).
Thus the ratio of the decoupling effect to the total change of the effective coupling 
 \be \frac{\beta_{\lambda}^{(2)} \ln(M_+^{}/M_-^{})}{\beta_{\lambda}^{(1)} \ln (M_{pl}^{}/100\text{GeV})}\sim 
 \frac{\beta_{\lambda}^{(2)}}{\beta_{\lambda}^{(1)}} \frac{\ln \kappa^{-1}}{72},
 \label{logratio}
 \e
becomes sizable $\sim {\cal{O}}(0.1)$
if $\lambda_\varphi, \ \lambda_h \sim {\cal O}(0.1)$ and $\kappa\sim 10^{-3}$.

%
\begin{figure}
\begin{center}
\begin{tabular}{c}
\begin{minipage}{0.5\hsize}
\begin{center}
\includegraphics[width=8.5cm]{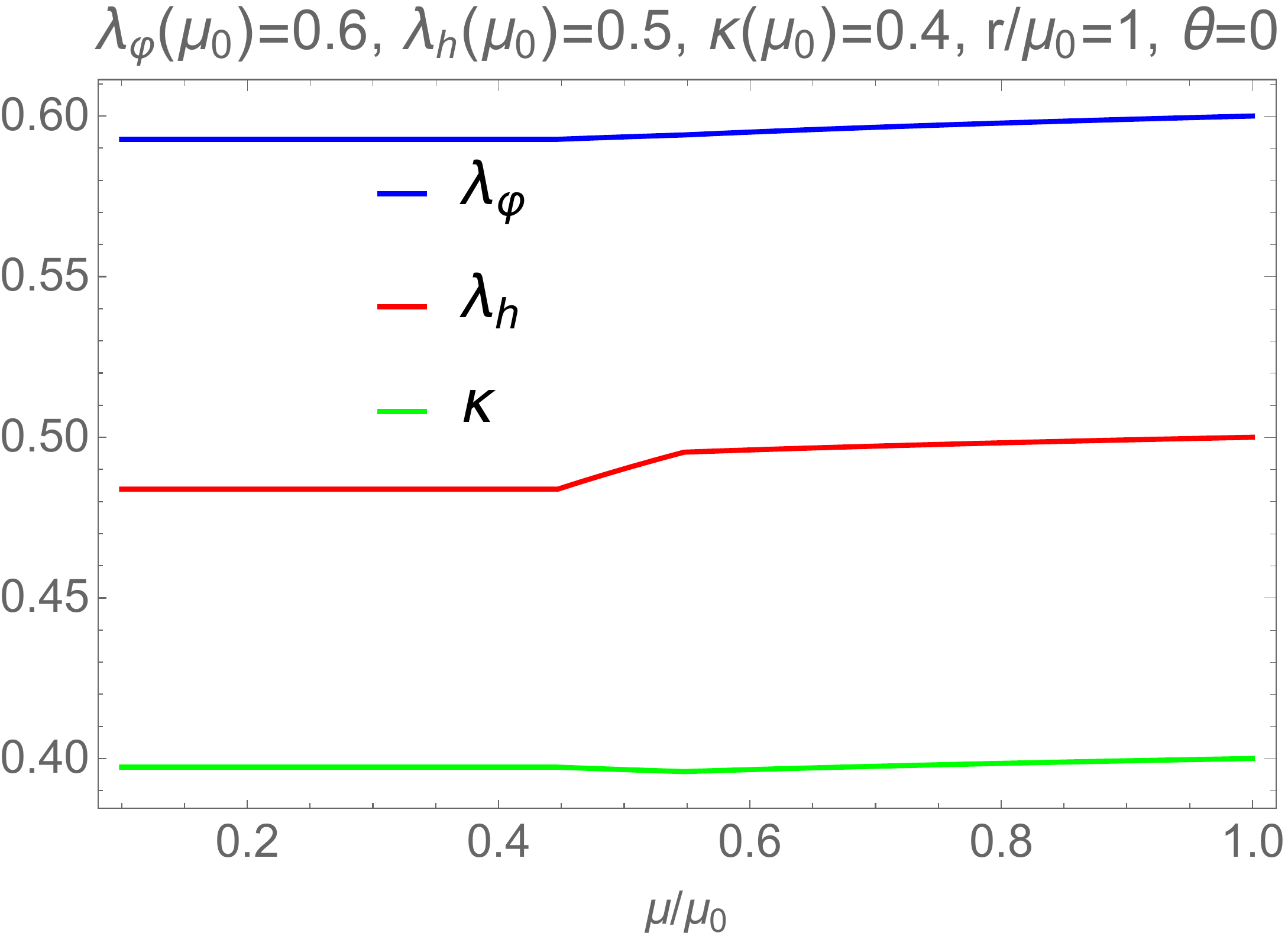}
\end{center}
\end{minipage}
\begin{minipage}{0.5\hsize}
\begin{center}
\includegraphics[width=8.5cm]{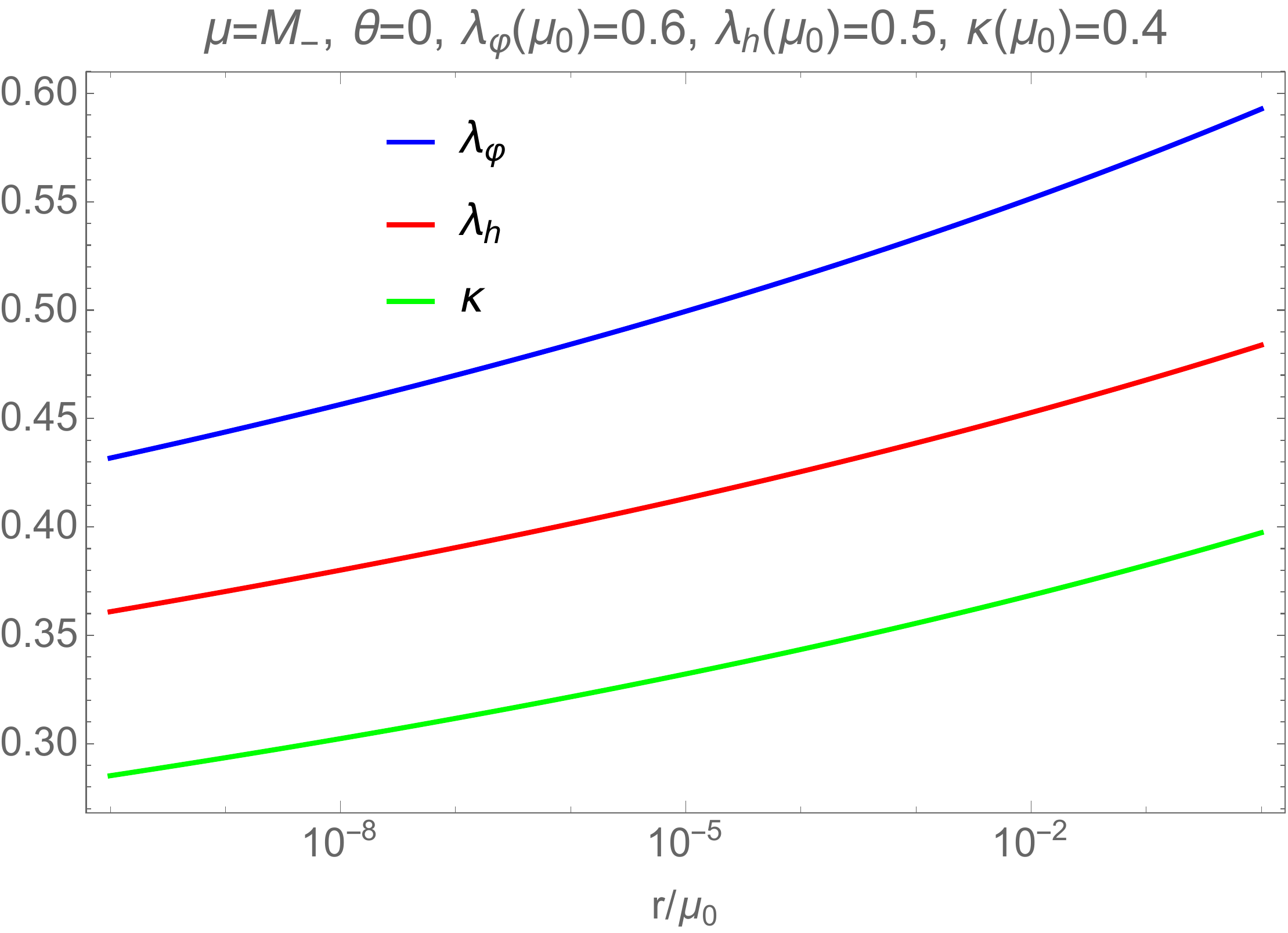}
\end{center}
\end{minipage}
\\
\\
\begin{minipage}{0.5\hsize}
\begin{center}
\includegraphics[width=8.5cm]{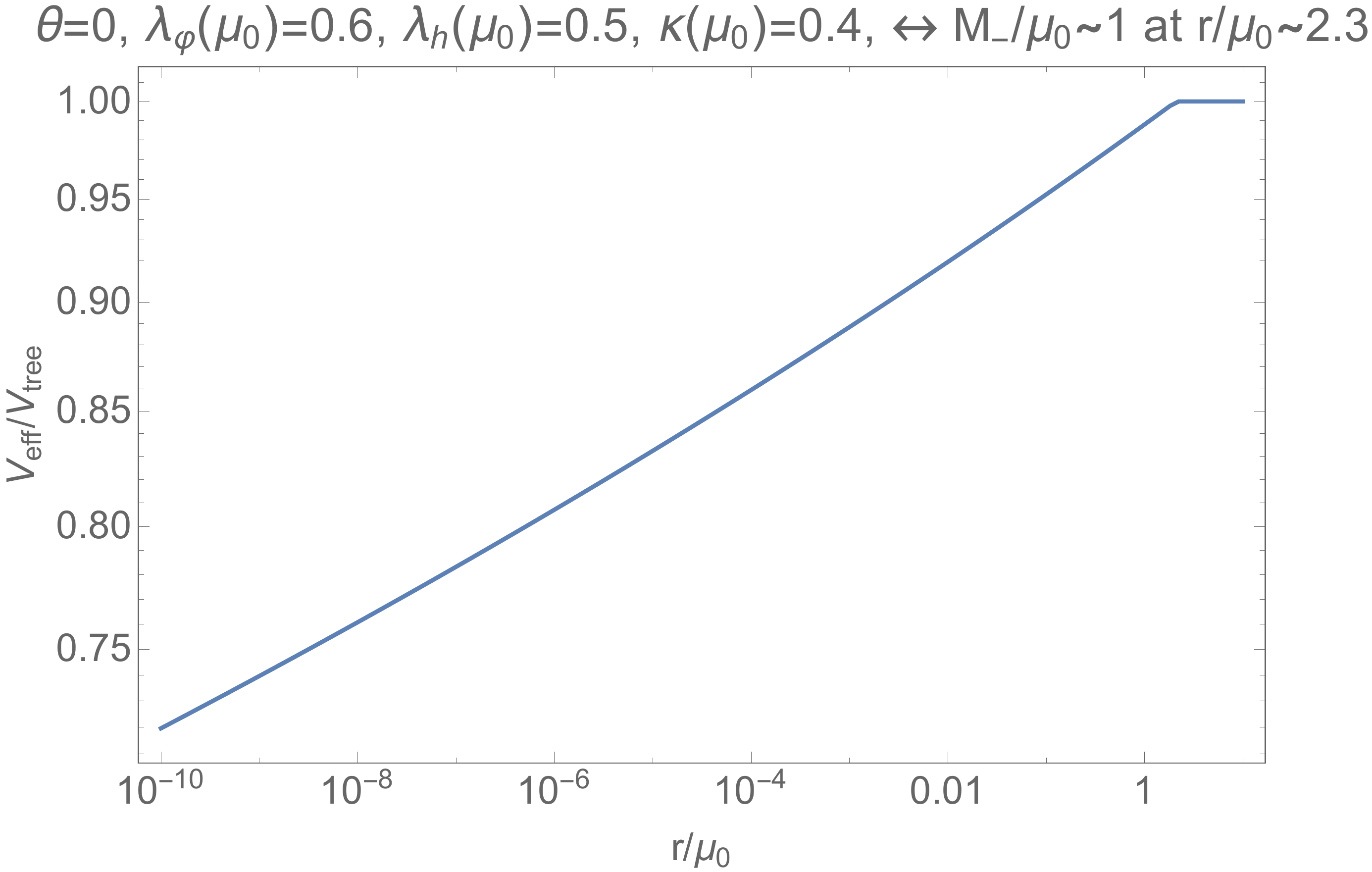}
\end{center}
\end{minipage}
\end{tabular}
\end{center}
\caption{Upper Left: The running coupling constants as functions of $\mu$ for $\theta=0$. Upper Right: The running coupling constants as a function of $r$ where we put $\tilde\mu=\text{min}(M_+^{},M_-^{})=M_-^{}$. Lower: The effective potential for $\theta=0$ as a function of $r$. 
}
\label{fig:run1}
\end{figure}
\begin{figure}
\begin{minipage}{0.5\hsize}
\begin{center}
\includegraphics[width=8cm]{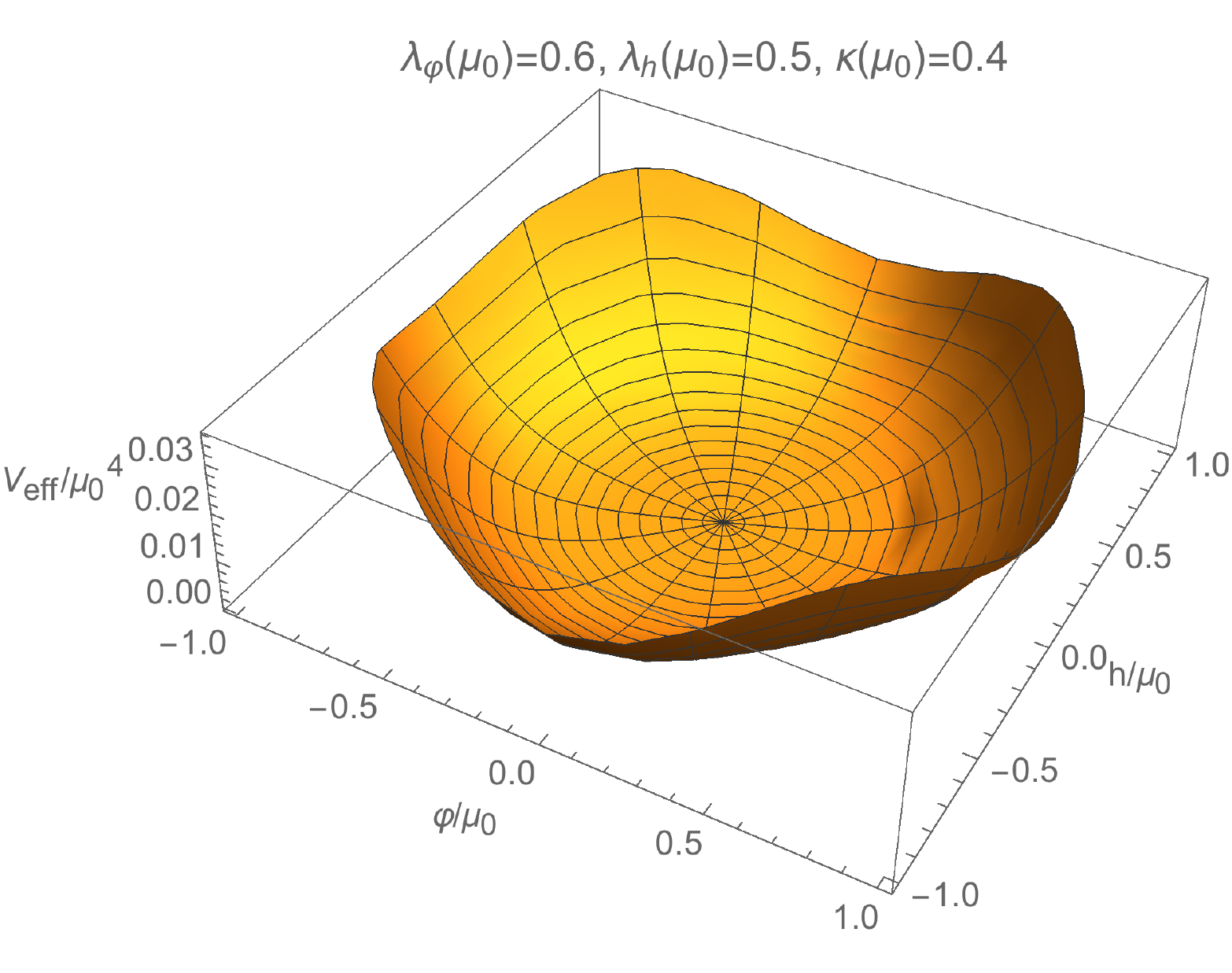}
\end{center}
\end{minipage}
\begin{minipage}{0.5\hsize}
\begin{center}
\includegraphics[width=8cm]{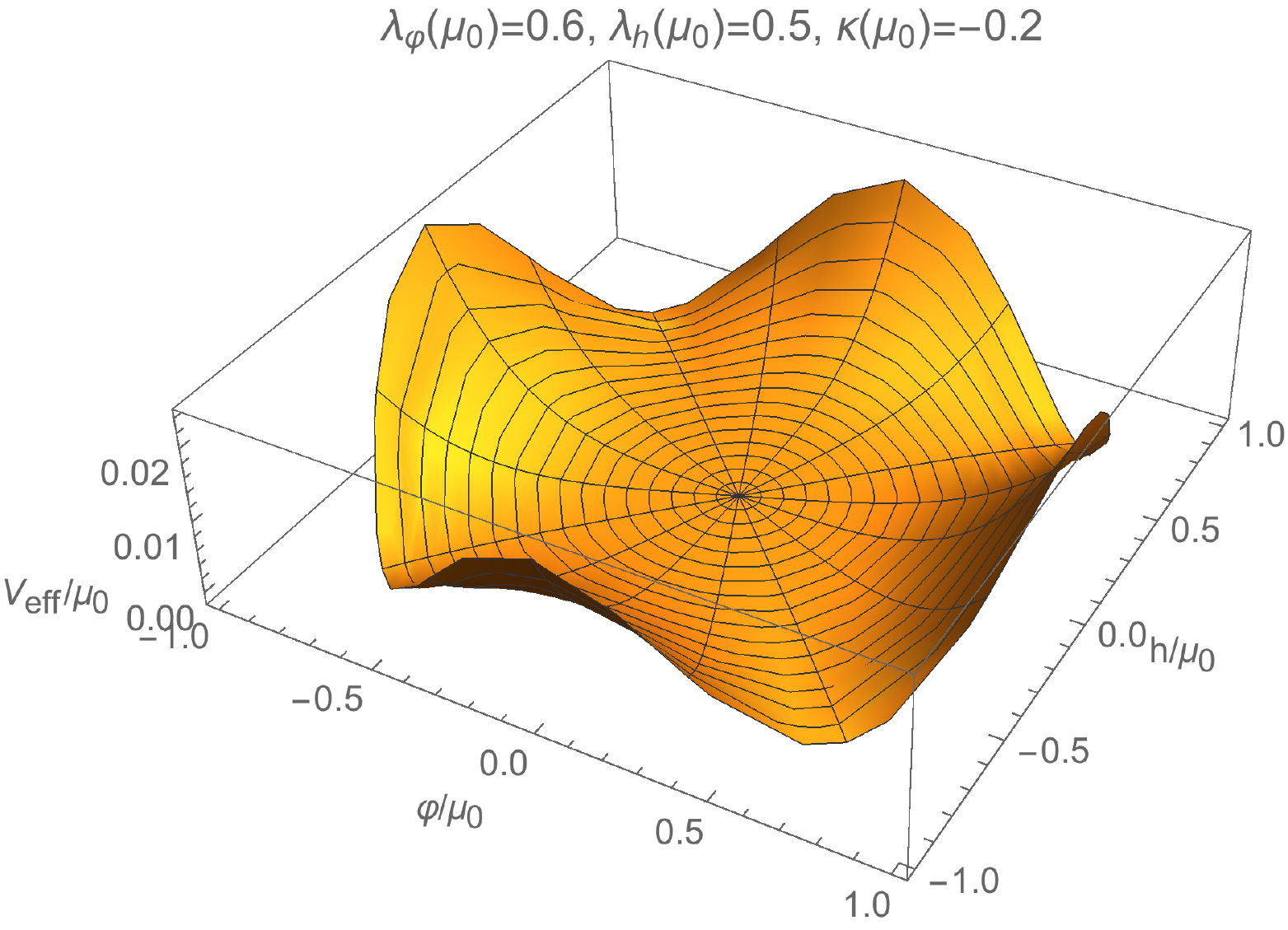}
\end{center}
\end{minipage}
\caption{Improved one-loop effective potentials of two real scalar model. The left (right) figure corresponds to the $\kappa>0\ (<0)$ case. 
} 
\label{fig:3dpotential}
\end{figure}

%
In the upper left panel of Fig.\ref{fig:run1}, we show the running couplings obtained from Eqs.(\ref{eq: scalar beta 1})-(\ref{eq: scalar beta 3}) in the direction $\theta=0$.  
Here, all the dimensional quantities are normalized by the initial value of the initial renormalization scale $\mu_0^{}$. 
We can 
see the step behaviors, particularly in $\lambda_{h}$, 
corresponding to the step functions in the beta functions. 
In the upper right panel of Fig.\ref{fig:run1}, we show the running couplings as a function of $r$ by putting $\tilde \mu=\text{min}(M_+^{},M_-^{})=M_-^{}$. 
Note that, when $r$ becomes sufficiently large so that 
$M_-^{}$ is larger than $\mu_0^{}$, all the couplings stop running because all the particles are decoupled. 
By using these running couplings,  the improved effective potential is given by
\aln{ V(r,\theta)&=\frac{\lambda_\varphi^{}}{4!}\varphi^4+\frac{\kappa}{4}\varphi^2h^2+\frac{\lambda_h^{}}{4!}h^4\bigg|_{\tilde{\mu}=M_-^{}}^{}\nonumber
\\
&=r^4\left(\frac{\lambda_\varphi^{}}{4!}\cos^2 \theta+\frac{\kappa}{4}\cos \theta \sin \theta+\frac{\lambda_h^{}}{4!}\sin^2
\theta\right)\bigg|_{\tilde{\mu}=M_-^{}(r,\theta)}^{}.
\label{eq:two scalar improvement}
}
In the lower panel in Fig.\ref{fig:run1}, we show a result of our numerical calculation. Here, $\theta$ is fixed to be zero and the potential is normalized by its tree-level one without any improvement. As we mentioned before, this ratio becomes one  
in the region $M_-^{}\geq \mu_0^{}$ because all the particles decouple and there is no running effects.    
By changing $r$ and $\theta$, one can obtain the improved effective potential in the whole region of $(\varphi,h)$, and it is shown in Fig.\ref{fig:3dpotential}. Here, the left  (right) panel corresponds to the $\kappa>0\ (<0)$ case. 
%

\section{Summary and Discussions}\label{sec:summary}
In this paper, we studied  RG improvement of the effective action of a Higgs-Yukawa model and 
a two real scalar model based on the decoupling approach \cite{Casas:1998cf}. 
In this approach, the RGEs can automatically 
take into account decouplings of massive particles and their threshold corrections 
by introducing appropriately chosen step functions.
In models with multiple scalar fields, the mass matrix
is  a function of field values, 
and decoupling scales of  heavy particles  depend on them. 
As a result, we need to use a different RGE 
on each point of the field space.  
Furthermore, because of the scalar mixing, the effective potential generally has non-polynomial terms.
By expanding them, we also get field-dependent $\beta$-functions. 
Taking these effects into account, we have obtained the RG improved effective potential for the 
two real scalar model. If there is a large hierarchy between multiple mass scales, the effect
might be sizable. 
We also investigated  decoupling scales when various fields with different masses simultaneously
exchange in a loop diagram. A simple example is the one-loop wave function renormalization 
in the Higgs-Yukawa model. 
Since both of heavy and light fields contribute to the diagrams, the decoupling scales are
given by neither of their masses but complicated combinations of them.

We now comment on an extension of the method to higher loop corrections. 
In \cite{Casas:1998cf}, the authors proposed a generalization to 2-loop diagrams
and showed that the decoupling scales are well-behaved
in potentially dangerous limits where multiple mass scales have infinitely large hierarchy $M_i^{}/M_j^{} \rightarrow 0$
or $\infty$.  
A key identity here is 
$f(x){\cal{D}}\theta(f(x))=f(x)\delta(f(x)){\cal{D}}f(x)\equiv 0$, where ${\cal D}$ is the operator 
defined in Eq.(\ref{eq: RGE of effective action}). 
Suppose that the model has a loop expansion parameter $g^2 \hbar$. 
Then the effective action is written up to $n$-th loop as
\begin{equation}
\Gamma = S^{(0)}+   \sum_{k=1}^{n} (g^2 \hbar)^k  S^{(k)} ,
\end{equation}
where $S^{(0)}$ is the classical action and $S^{(k)}$ are $k$-th loop contributions to the effective action which
can be further decomposed as 
$
S^{(k)} =  \sum_{p=0}^{k} (\log \mu )^p \sum_{\cal O} S_{\cal O}^{k,k-p} .
$
Here ${\cal O}$ denotes both of differences of operators, namely $(\partial\varphi)^2, \varphi^2, \varphi^4$ etc. 
and  differences of Feynman diagrams. $p=1$ terms determines the beta and gamma functions. 
Namely the invariance of the effective action $\Gamma$  under
a change of renormalization scale $\mu$  relates $(g\hbar)^k (\log \mu)^p$ terms 
for $p>1$ with the  $p=1$ terms. 
In order to define the effective action with  step functions as in (\ref{eq:generalization}),
it is necessary to respect the relations between various different terms. 
One possibility will be to introduce a different step function in each diagram with $p=1$
as we did in the Higgs-Yukawa model for one-loop calculations, and determine
the step functions for $p>1$ terms so as to respect the RG relations, but 
since the logarithmic structures of higher loop diagrams are complicated, 
more detailed studies will be necessary. We leave it for future investigations. 
This prescription is slightly different from the 2-loop prescription proposed by  \cite{Casas:1998cf}.
In their prescription, logarithmic terms are expanded with respect to $1/M$ where $M$ is 
the largest mass scale in the model. 
Here the question is whether we can reduce
various logarithms including multiple mass scales such as $\log[(z M^2+(1-z)m^2)/\mu^2]$ 
into those with a single mass, $\log(M^2/\mu^2)$. 
As we saw in the Higgs-Yukawa model, in order to avoid divergences of Feynman parameter 
integrals, we need to treat logarithmic factors in loop integrals without using an expansion with respect to $1/M$.
Thus the decoupling scales are not simply given by a single heavy mass, but a complicated combination
of various masses.
In the case of $\mu_G$ and $\mu_{\tilde{G}}$ in Eq.(\ref{stepfunctions4}) of the Higgs-Yukawa model, 
they are in between the heavy and the light mass scales
and monotonically increase as the heavy mass becomes larger as in Fig.\ref{fig:decoupling scale}.
They are also well-behaved in the potentially dangerous limit; $A=M_\psi/M_\phi \rightarrow 0, \infty$.
These behaviors are also expected to hold even for higher loop effects. 
We want to come back to these problems in future.

\section*{Acknowledgements} 
This work of SI is supported in part by Grants-in-Aid for Scientific
Research (No.\ 16K05329) from the Japan Society for the Promotion of Science.
The work of KK is supported by the Grant-in-Aid for JSPS Research Fellow, Grant Number 17J03848. 

\appendix 
\def\thesection{Appendix \Alph{section}}
\section*{Appendix: One-loop calculations of Higgs-Yukawa model}
In this appendix, we give the derivation of Eq.(\ref{eq: one-loop effective action}). 
First, let us consider the effective action in the bosonic background;
\aln{\Gamma_{1\text{loop}}^B[\phi]=\frac{i}{2}\text{tr}\log\left(\Box+M_{\phi}^{}(\phi_{cl}^{})^2\right)-i\text{tr}\log\left(i\cancel{\partial}-M_{\psi}^{}(\phi)\right)
}
The first term of the right hand side corresponds to the scalar contributions to the one-loop effective potential.
The second term of the fermionic contributions can be rewritten as
\aln{\text{tr}\log\left(i\cancel{\partial}-M_{\psi}^{}(\phi)\right)
&=\frac{1}{2}\left[\text{tr}\log\left(i\cancel{\partial}-M_{\psi}^{}(\phi)\right)+\text{tr}\log\left(-i\cancel{\partial}-M_{\psi}^{}(\phi)\right)\right]
\nn
&=\frac{1}{2}\text{tr}\log\left(\cancel{\partial}^2+M_\psi^{}(\phi)^2-ig(\cancel{\partial}\phi)\right)
\nn
&=\int d^4x\left[iV_{1\text{loop}}^{(F)}(\phi)+\frac{1}{2}\int \frac{d^dq}{(2\pi)^d}\text{tr}^{(\gamma)}\log\left(1+\frac{ig(\cancel{\partial}\phi)}{q^2-M_\psi^{}(\phi)^2}\right)\right],
\label{eq:one loop boson part}
}
where we have used the fact that the action is invariant under $(x^\mu,\gamma^\mu)\rightarrow (-x^\mu,-\gamma^\mu)$. Then, by expanding the second term of Eq.(\ref{eq:one loop boson part}) with respect to $g(\cancel{\partial}\phi)/(q^2-M_\psi^{}(\phi)^2)$, we obtain 
\aln{ &-\frac{(ig)^2\mu^\epsilon}{4}\int d^dx\int \frac{d^dq}{(2\pi)^d}\text{tr}^{(\gamma)}\frac{(\cancel{\partial}\phi)^2}{(q^2-M_\psi^{}(\phi)^2)^2} \nn
& =\frac{ig^2}{16\pi^2}\int d^4x(\partial_\mu^{}\phi)^2\int_0^1 dz\left(\frac{2}{\epsilon}-\gamma+\log4\pi-\log\frac{M_\psi^{}(\phi)^2}{\mu^2}\right)
\nn
&  
\underset{\overline{\text{MS}}}{\longrightarrow}
-\frac{ig^2}{16\pi^2}\int d^4x(\partial_\mu^{}\phi)^2\log\left(\frac{M_\psi^{}(\phi)^2}{\mu^2}\right),
}
where $\epsilon=4-d$. This term gives the wave function renormalization in the RGE. 

Next, let us consider the fermionic background, i.e. the third term of the right hand side of Eq.(\ref{eq: one loop Higgs Yukawa}). For our present purpose, it is sufficient to consider a constant $\phi$. 
By expanding the logarithm, we have
\aln{&\frac{i}{2}\frac{\int {\cal{D}}\delta \phi\int {\cal{D}} \delta\psi\int {\cal{D}} \overline{\delta\psi}e^{i\delta S_0^{}}
\delta S_{int}^{}\delta S_{int}^{}}{Z_0^{}[\phi_{cl}^{}]} \nn
& =ig^2\mu^\epsilon\int d^4x\int d^4y\int \frac{d^dq}{(2\pi)^d}\int \frac{d^dk}{(2\pi)^d}\frac{ie^{-i(q+k)(x-y)}}{q^2-M_\phi^{}(\phi)^2}
 \frac{\overline{\psi}(x) i(\cancel{k}+M_\psi^{}(\phi)) \psi(y)}{k^2-M_\psi^{}(\phi)^2}.
}
Then, by introducing a relative coordinate $z\equiv y-x$ and expanding $\psi(x+z)$ with respect to $z$, we obtain 
\aln{
&-ig^2\mu^\epsilon\int d^4x\bigg[\int \frac{d^dq}{(2\pi)^d}\frac{\overline{\psi}(x)(-\cancel{q}+M_\psi^{}(\phi))\psi(x)}{(q^2-M_\phi^{}(\phi)^2)(q^2-M_\psi^{}(\phi)^2)}
\nn
&\h{1cm}+i\int \frac{d^dq}{(2\pi)^d}\frac{d}{dq_\mu^{}}\left(\frac{1}{q^2-M_\phi^{}(\phi)^2}\right)
\overline{\psi}(x)\frac{-\cancel{q}+M_\psi^{}(\phi)}{q^2-M_\psi^{}(\phi)^2}\partial_\mu^{}\psi(x)+\cdots
\bigg],
\label{eq: fermion calculation}
}
where we have used the following identity:
\be \int \frac{d^dz}{(2\pi)^d}z^\mu e^{-i(q+k)z}=i\frac{\partial}{\partial q^\mu}\delta^{(d)}(q+k) =-i\delta^{(d)}(q+k)\frac{\partial}{\partial q^\mu}.
\e
The first term in Eq.(\ref{eq: fermion calculation}) gives the one-loop correction to the Yukawa term:
\aln{&-ig^2\mu^\epsilon\int d^4x\int \frac{d^dq}{(2\pi)^d}\frac{\overline{\psi}(x)(-\cancel{q}+M_\psi^{}(\phi))\psi(x)}{(q^2-M_\phi^{}(\phi)^2)(q^2-M_\psi^{}(\phi)^2)}
\nn
&=\frac{g^2}{(16\pi^2)}\int d^4xM_\psi^{}(\phi)\overline{\psi}\psi\int_0^1dz\left(\frac{2}{\epsilon}-\gamma+\log4\pi-\log\left(\frac{zM_\phi^{}(\phi)^2+(1-z)M_\psi^{}(\phi)^2}{\mu^2}\right)\right)
\nn
&\underset{\overline{\text{MS}}}{\longrightarrow}-\frac{g^2}{(16\pi^2)}\int d^4xM_\psi^{}(\phi)\overline{\psi}\psi\int_0^1dz\log\left(\frac{zM_\phi^{}(\phi)^2+(1-z)M_\psi^{}(\phi)^2}{\mu^2}\right).
}
The second term of Eq.(\ref{eq: fermion calculation}) gives the one-loop correction to the kinetic term of $\psi$:
\aln{&g^2\mu^\epsilon\int d^4x\int \frac{d^dq}{(2\pi)^d}\frac{2q^\mu}{(q^2-M_\phi^{}(\phi)^2)^2}
\overline{\psi}(x)\frac{-\cancel{q}+M_\psi^{}(\phi)}{q^2-M_\psi^{}(\phi)^2}\partial_\mu^{}\psi(x)
\nn
&=-\frac{2g^2\mu^\epsilon}{d}\int d^4x\int_0^1 dz\int_0^{z}dy\int \frac{d^dq}{(2\pi)^d}\frac{2!q^2}{[q^2-zM_\phi^{}(\phi)-(1-z)M_\psi^{}(\phi)]^3}\overline{\psi}\cancel{\partial}\psi
\nn
&=-\frac{g^2}{16\pi^2}\int d^4x\overline{\psi}i\cancel{\partial}\psi\int_0^1 dzz\left(\frac{2}{\epsilon}-\gamma+\log4\pi-\log\left(\frac{zM_\phi^{}(\phi)^2+(1-z)M_\psi^{}(\phi)^2}{\mu^2}\right)\right)
\nn
&\underset{\overline{\text{MS}}}{\longrightarrow}\frac{g^2}{16\pi^2}\int d^4x\overline{\psi}i\cancel{\partial}\psi\int_0^1 dzz\log\left(\frac{zM_\phi^{}(\phi)^2+(1-z)M_\psi^{}(\phi)^2}{\mu^2}\right).
}


\begin{thebibliography}{40}

\bibitem{Bando:1992wy} 
  M.~Bando, T.~Kugo, N.~Maekawa and H.~Nakano,
  ``Improving the effective potential: Multimass scale case,''
  Prog.\ Theor.\ Phys.\  {\bf 90}, 405 (1993)
  doi:10.1143/PTP.90.405, 10.1143/ptp/90.2.405
  [hep-ph/9210229].

\bibitem{Casas:1998cf} 
  J.~A.~Casas, V.~Di Clemente and M.~Quiros,
  ``The Effective potential in the presence of several mass scales,''
  Nucl.\ Phys.\ B {\bf 553}, 511 (1999)
  doi:10.1016/S0550-3213(99)00262-X
  [hep-ph/9809275].

\bibitem{Holthausen:2011aa} 
  M.~Holthausen, K.~S.~Lim and M.~Lindner,
  ``Planck scale Boundary Conditions and the Higgs Mass,''
  JHEP {\bf 1202}, 037 (2012)
  doi:10.1007/JHEP02(2012)037
  [arXiv:1112.2415 [hep-ph]].

\bibitem{Bezrukov:2012sa} 
  F.~Bezrukov, M.~Y.~Kalmykov, B.~A.~Kniehl and M.~Shaposhnikov,
  ``Higgs Boson Mass and New Physics,''
  JHEP {\bf 1210}, 140 (2012)
  doi:10.1007/JHEP10(2012)140
  [arXiv:1205.2893 [hep-ph]].
  
\bibitem{Degrassi:2012ry} 
  G.~Degrassi, S.~Di Vita, J.~Elias-Miro, J.~R.~Espinosa, G.~F.~Giudice, G.~Isidori and A.~Strumia,
  ``Higgs mass and vacuum stability in the Standard Model at NNLO,''
  JHEP {\bf 1208}, 098 (2012)
  doi:10.1007/JHEP08(2012)098
  [arXiv:1205.6497 [hep-ph]].

\bibitem{Iso:2012jn} 
  S.~Iso and Y.~Orikasa,
  ``TeV Scale B-L model with a flat Higgs potential at the Planck scale - in view of the hierarchy problem -,''
  PTEP {\bf 2013}, 023B08 (2013)
  doi:10.1093/ptep/pts099
  [arXiv:1210.2848 [hep-ph]].
  
\bibitem{Buttazzo:2013uya} 
  D.~Buttazzo, G.~Degrassi, P.~P.~Giardino, G.~F.~Giudice, F.~Sala, A.~Salvio and A.~Strumia,
  ``Investigating the near-criticality of the Higgs boson,''
  JHEP {\bf 1312}, 089 (2013)
  doi:10.1007/JHEP12(2013)089
  [arXiv:1307.3536 [hep-ph]].
    
  
\bibitem{Kawana:2015tka} 
  K.~Kawana,
  ``Criticality and inflation of the gauged B – L model,''
  PTEP {\bf 2015}, 073B04 (2015)
  doi:10.1093/ptep/ptv093
  [arXiv:1501.04482 [hep-ph]].
  
\bibitem{Hamada:2015fma} 
  Y.~Hamada and K.~Kawana,
  ``Vanishing Higgs Potential in Minimal Dark Matter Models,''
  Phys.\ Lett.\ B {\bf 751}, 164 (2015)
  doi:10.1016/j.physletb.2015.10.006
  [arXiv:1506.06553 [hep-ph]].
  
\bibitem{Alekhin:2012py} 
  S.~Alekhin, A.~Djouadi and S.~Moch,
  ``The top quark and Higgs boson masses and the stability of the electroweak vacuum,''
  Phys.\ Lett.\ B {\bf 716}, 214 (2012)
  doi:10.1016/j.physletb.2012.08.024
  [arXiv:1207.0980 [hep-ph]].
  
 
\bibitem{Moch:2014tta} 
  S.~Moch {\it et al.},
  ``High precision fundamental constants at the TeV scale,''
  arXiv:1405.4781 [hep-ph].

\bibitem{Cortiana:2015rca} 
  G.~Cortiana,
  ``Top-quark mass measurements: review and perspectives,''
  Rev.\ Phys.\  {\bf 1}, 60 (2016)
  doi:10.1016/j.revip.2016.04.001
  [arXiv:1510.04483 [hep-ex]].

\bibitem{Coleman:1973jx} 
  S.~R.~Coleman and E.~J.~Weinberg,
  ``Radiative Corrections as the Origin of Spontaneous Symmetry Breaking,''
  Phys.\ Rev.\ D {\bf 7}, 1888 (1973).
  doi:10.1103/PhysRevD.7.1888

\bibitem{Einhorn:1983fc} 
  M.~B.~Einhorn and D.~R.~T.~Jones,
  ``A New Renormalization Group Approach To Multiscale Problems,''
  Nucl.\ Phys.\ B {\bf 230}, 261 (1984).
  doi:10.1016/0550-3213(84)90127-5

\bibitem{Ford:1996yc} 
  C.~Ford and C.~Wiesendanger,
  ``Multiscale renormalization,''
  Phys.\ Lett.\ B {\bf 398}, 342 (1997)
  doi:10.1016/S0370-2693(97)00237-2
  [hep-th/9612193].
  
\bibitem{Steele:2014dsa} 
  T.~G.~Steele, Z.~W.~Wang and D.~G.~C.~McKeon,
  ``Multiscale renormalization group methods for effective potentials with multiple scalar fields,''
  Phys.\ Rev.\ D {\bf 90}, no. 10, 105012 (2014)
  doi:10.1103/PhysRevD.90.105012
  [arXiv:1409.3489 [hep-ph]].

\bibitem{Bando:1992np} 
  M.~Bando, T.~Kugo, N.~Maekawa and H.~Nakano,
  ``Improving the effective potential,''
  Phys.\ Lett.\ B {\bf 301}, 83 (1993)
  doi:10.1016/0370-2693(93)90725-W
  [hep-ph/9210228].


\bibitem{Appelquist:1974tg} 
  T.~Appelquist and J.~Carazzone,
  ``Infrared Singularities and Massive Fields,''
  Phys.\ Rev.\ D {\bf 11}, 2856 (1975).
  doi:10.1103/PhysRevD.11.2856
  
\bibitem{Symanzik:1973vg} 
  K.~Symanzik,
  ``Infrared singularities and small distance behavior analysis,''
  Commun.\ Math.\ Phys.\  {\bf 34}, 7 (1973).
  doi:10.1007/BF01646540





\end{thebibliography}
\end{document}